\documentclass[twoside,12pt]{article}
\usepackage{epsfig}

\def\Journal#1#2#3#4{{#1} {#2} (#4) #3 }

\def\NPA{{\em Nucl. Phys.} A}

\def\NPB{{\em Nucl. Phys.} B}

\def\PLB{{\em Phys. Lett.} B}

\def\PRL{\em Phys. Rev. Lett.}

\def\PREP{\em Phys. Rep.}

\def\PRD{{\em Phys. Rev.} D}
\def\PRC{{\em Phys. Rev.} C}

\def\EPJA{{\em Eur. Phys. J.} A}

\def\RMP{{\em Rev. Mod. Phys.}}

\newcommand{\be}{\begin{equation}}
\newcommand{\ee}{\end{equation}}
\newcommand{\bea}{\begin{eqnarray}}
\newcommand{\eea}{\end{eqnarray}}

\begin{document}
\title{ \vspace{1cm} Nuclear Chiral Dynamics and Phases of QCD}
\author{W. Weise \\
\\
Physik-Department, Technische Universit\"at M\"unchen,\\ D-85747 Garching, Germany}
\maketitle

\begin{abstract} This presentation starts with a brief review of our current 
picture of QCD phases, derived from lattice QCD thermodynamics and from
models based on the symmetries and symmetry breaking patterns of QCD. 
Typical approaches widely used in this context are the PNJL and chiral quark-meson
models. It is pointed out, however, that the modeling of the phase diagram in terms of 
quarks as quasiparticles misses important and well known nuclear physics constraints. 
In the hadronic phase of QCD governed by confinement and spontaneously 
broken chiral symmetry, in-medium chiral effective field theory is the appropriate framework,
with pions and nucleons as active degrees of freedom.  Nuclear chiral thermodynamics is
outlined and the liquid-gas phase transition is described. The density and temperature
dependence of the chiral condensate is deduced. As a consequence of two- and three-body 
correlations in the nuclear medium, no tendency towards a first-order chiral phase
transition is found at least up to twice the baryon density of normal nuclear matter and 
up to temperatures of about 100 MeV. Isospin-asymmetric nuclear matter and neutron matter
are also discussed. An outlook is given on new tightened constraints for the equation-of-state 
of cold and highly compressed matter as implied by a recently observed two-solar-mass
neutron star.
\end{abstract}

\section{Introduction\\ QCD phase diagram: visions and facts}
\begin{figure}
\begin{center}
\includegraphics*[totalheight=9cm]{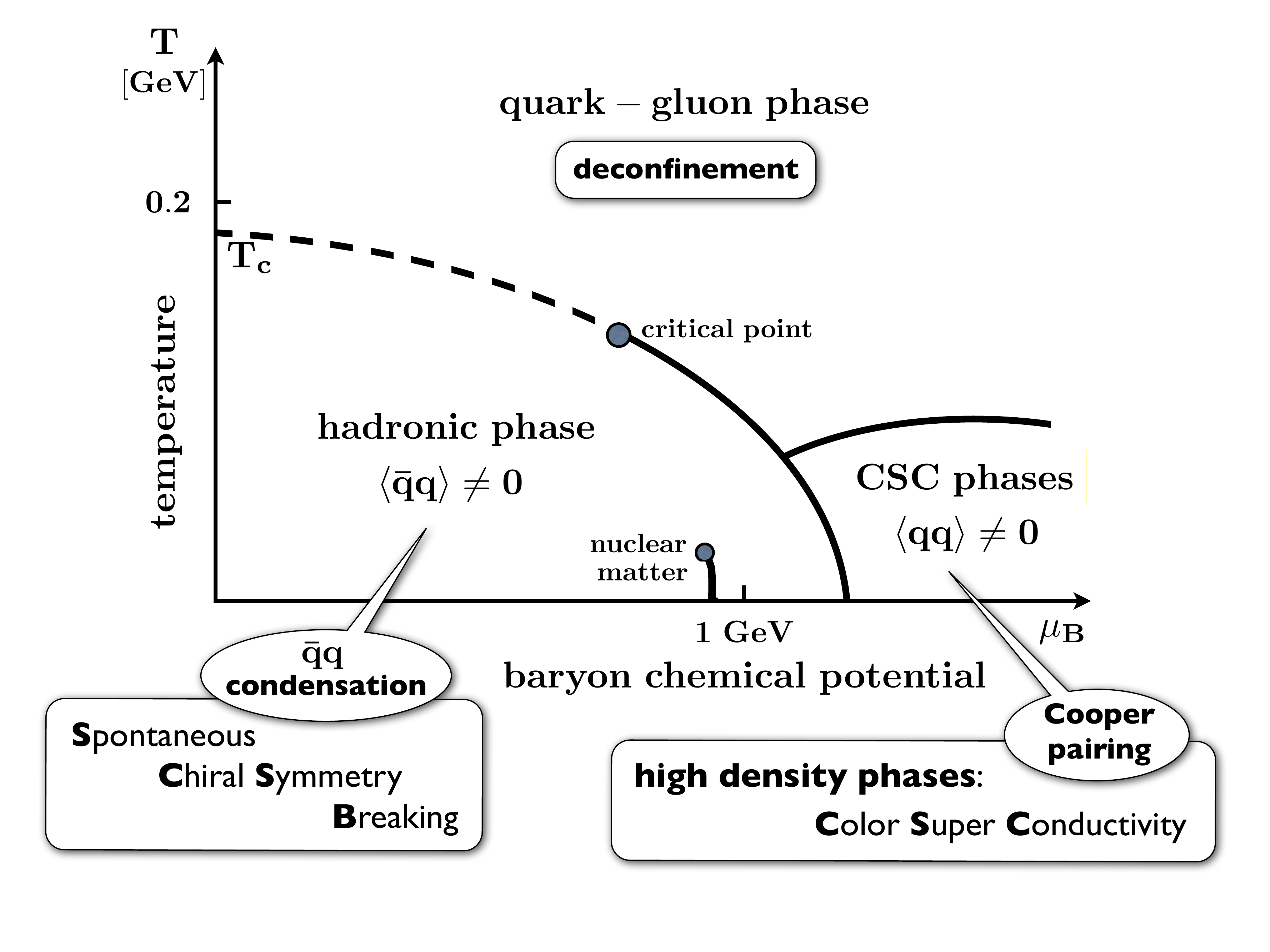}
\caption{Schematic picture of the QCD phase diagram in terms of temperature $T$ and baryon chemical
potential $\mu_B$,  sketching the deconfined quark-gluon phase at high temperature, the low-temperature
and low-density hadronic phase with spontaneously broken chiral symmetry, and the domain of high-density
(color-superconducting - CSC) phases featuring various forms of quark Cooper pairing. The dashed line
illustrates a chiral crossover transition ending at a hypothetical critical point where a first-order
phase transition boundary (solid line) begins. The first-order liquid-gas phase transition of nuclear
matter is also indicated.}
\label{Fig1}
\end{center}
\end{figure}

A theorist's sketch of the QCD phase diagram, plotted in the plane of temperature $T$
and baryon chemical potential $\mu_B$, usually looks like the one shown in Fig.\ref{Fig1}.
The hadronic phase, with quarks and gluons confined in mesons and baryons,
is thought to be separated from the deconfined quark-gluon phase by a crossover transition at 
low baryon chemical potential $\mu_B$ and at temperatures $T$ of order $\Lambda_{QCD} \sim 0.2$ GeV. At
a critical point this crossover supposedly ends and turns, at larger values of the baryon chemical potential, into a first-order
phase transition. When the baryon chemical potential is further increased, various low-temperature 
color-superconducting phases are expected to emerge. 

At zero chemical potential, lattice simulations of full QCD thermodynamics with three quark flavors \cite{Ch2008, Bo2010, BP2010, Aoki2006} do indeed suggest a crossover scenario for the temperature dependence of the chiral
(quark) condensate, an order parameter spontaneously broken chiral symmetry, with 
a transition temperature $T_c$ around $170$ MeV. Lattice QCD extrapolations to non-zero (real) chemical potentials are notoriously difficult, however. Extensions of the phase diagram to baryonic matter and various forms of quark matter rely so far entirely on models. Only at asymptotically large quark chemical potentials the picture becomes simple again. At Fermi momenta of the order of several GeV, the quarks at the Fermi surface can be treated
using perturbative QCD. The gluon exchange quark-quark interaction is weakly attractive in color-antitriplett pairs.
Such a weakly attractive interaction is sufficient to form Cooper-pair-like diquarks in various combinations so that
color superconducting phases emerge \cite{Al2008}.  However, the more quantitative density scale at which such transitions take place cannot be predicted. 

Basically all models of the QCD phase diagram (briefly summarized in Section 2) work with dynamical quarks as quasiparticles and do not take into account baryons and their correlations. An apparently generic result of such models is the fact that the first order transition line at low temperature meets the chemical potential axis at a quark chemical potential around $\mu_q \sim 0.3$ GeV, the typical constituent quark mass scale, corresponding to a baryon chemical potential $\mu_B$ close to  the nucleon mass. This, however, is the terrain of nuclear physics with its enormously rich and well explored phenomenology. No sign of a first-order chiral or deconfinement  phase transition is visible in that area right in the center of the QCD phase diagram. The only known phase transition in nuclear matter is the one from a Fermi liquid to an interacting Fermi gas. This motivates the present essay as a proposal for taking the well-known nuclear physics constraints more seriously into account (see Section 3). This presentation is in parts an update of a previous review \cite{We2010}. It closes with a brief discussion of new constraints from neutron star observations on the equation of state at high density and low temperature. 

%
%

\newpage

\section{Modeling the QCD phase diagram}

Confinement and spontaneous chiral symmetry breaking are governed by  two basic symmetry principles of QCD:
\begin{itemize}
\item{The symmetry associated with the center $Z(3)$ of the local $SU(3)_c$ color gauge group is exact in the limit of pure gauge QCD, realized for {\it infinitely heavy} quarks. In the high-temperature, deconfinement phase of QCD this $Z(3)$ symmetry is spontaneously broken, with the Polyakov loop acting as the order parameter.}
\item{Chiral $SU(N_f)_R\times SU(N_f)_L$ symmetry is an exact global symmetry of QCD with $N_f$ {\it massless} quark flavors.  In the low-temperature (hadronic) phase this symmetry is spontaneously broken down to the flavor group $SU(N_f)_V$ (the isospin group for $N_f = 2$ and the ``eightfold way" for $N_f = 3$). As a consequence there exist $N_f^2 - 1$ pseudoscalar Nambu-Goldstone bosons and the QCD vacuum hosts a strong quark condensate.}
\end{itemize} 

These symmetries and symmetry breaking patterns serve as guiding principles for constructing models of the QCD phases. The basic question is about the intertwining of spontaneous chiral symmetry breaking with confinement.

\subsection{Chiral condensate and Polyakov loop}

The order parameter of spontaneously broken chiral symmetry is the quark condensate, $\langle \bar{q} q \rangle$.  The  disappearence of this condensate, by its melting above a characteristic transition temperature scale, signals the restoration of chiral symmetry in its unbroken Wigner-Weyl realization. The transition from confinement to deconfinement in QCD is likewise controlled by an order parameter, the Polyakov loop. A non-vanishing Polyakov loop $\Phi$ reflects the spontaneously broken $Z(3)$ symmetry characteristic of the deconfinement phase. The Polyakov loop vanishes in the low-temperature, confinement sector of QCD.

Two limiting cases are of interest in this context. In the pure gauge limit of QCD, corresponding to infinitely heavy quarks, the deconfinement transition is established as a first order phase transition with a critical temperature of about $270$ MeV. In the limit of massless $u$ and $d$ quarks, on the other hand, the isolated chiral transition appears as a second order phase transition at a significantly lower critical temperature. This statement is based on calculations using Nambu - Jona-Lasinio (NJL) type models \cite{NJL61, VW91, HK94} which incorporate the correct spontaneous chiral symmetry breaking mechanism but ignore confinement. The step from first or second order phase transitions to crossovers is understood as a consequence of explicit symmetry breaking. The $Z(3)$ symmetry is explicitly broken by the mere presence of quarks with non-infinite masses. Chiral symmetry is explicitly broken by non-zero quark masses. It is then a challenging question whether and how the chiral and deconfinement transitions get dynamically entangled in just such a way that they finally occur within overlapping transition temperature intervals.

\begin{figure}
\begin{center}
\includegraphics*[totalheight=8cm]{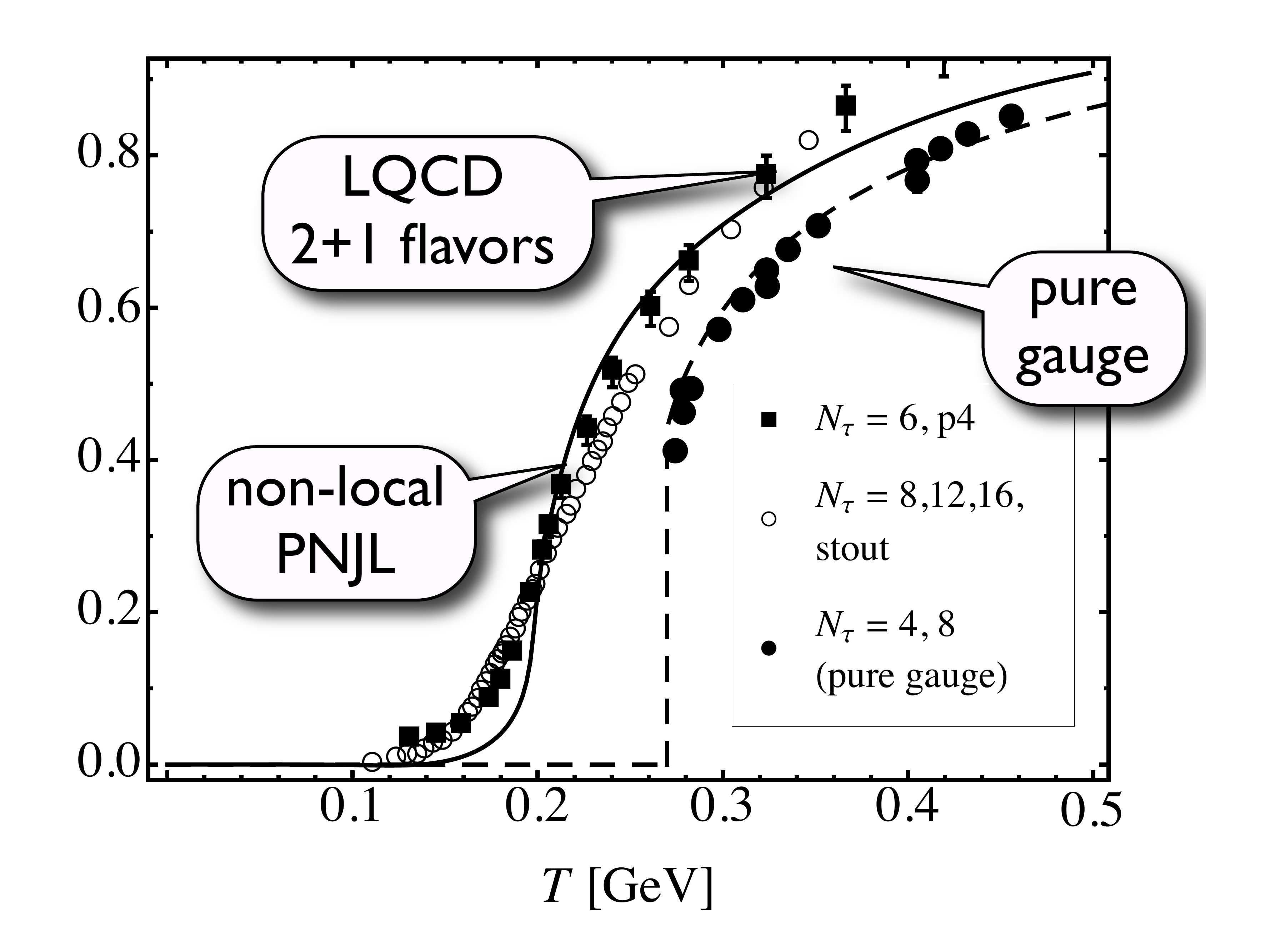}
\caption{Pure gauge and full QCD lattice results for the Polyakov loop as function of temperature \cite{Ch2008}. The solid line is calculated in an advanced (non-local) PNJL model \cite{HKW2011}, to be discussed later in the text.}
\label{Fig2}
\end{center}
\end{figure}

Confinement implies spontaneous chiral symmetry breaking, but the reverse is not necessarily true. There is no a priori reason why the chiral and deconfinement transitions should be intimitely connected. Nonetheless this appears to be the case in lattice QCD computations with almost physical quark masses. Results from lattice QCD thermodynamics \cite{Ch2008} with 2+1 flavors (at zero baryon chemical potential) gave a chiral crossover transition temperature of about 190 MeV.  Recent improved computations \cite{Bo2010, BP2010} indicate a shift of the chiral transition to somewhat lower temperatures, around 160 MeV, consistent with earlier lattice simulations \cite{Aoki2006}. Lattice QCD results for the temperature dependence of the Polyakov loop have remained remarkably stable over the years. Pure gauge QCD on the lattice, with full gluon dynamics but without quarks (or equivalently, with infinitely heavy quarks) has established a first-order deconfinement transition at a critical temperature $T_c \simeq 270$ MeV. Once light quarks are added, the $Z(3)$ center symmetry is explicitly broken by the presence of these quarks and the first-order deconfinement transition turns into a crossover, with a significantly reduced transition temperature around 200 MeV.  This is shown in Fig.\ref{Fig2} where Polyakov loop results from pure gauge lattice computations are compared to full QCD simulations with $N_f = 2+1$ quark flavors.  

%
%
 
\subsection{The non-local PNJL model}

Insights concerning the issue of a possible intertwining of chiral and deconfinement transitions can be gained from a model based on a minimal synthesis of the NJL-type spontaneous chiral symmetry breaking mechanism and confinement implemented through Polyakov loop dynamics. This PNJL model \cite{Fu2003,RTW2006} is specified by the following action:
\begin{eqnarray}
{\cal S} = \int_0^{\beta=1/T}d\tau \int_V d^3x\left[\psi^\dagger\partial_\tau\psi-{\cal H}(\psi,\psi^\dagger,\phi)\right] - {V\over T}\,{\cal U}(\Phi,T)~.
\end{eqnarray}
It introduces the Polyakov loop, 
\begin{equation}
\Phi = N_c^{-1}\,Tr\exp(i\phi/T)~,
\end{equation}
with a homogeneous temporal gauge field, $\phi = \phi_3\lambda_3 + \phi_8\lambda_8\in SU(3)$, coupled to the quarks. The dynamics of $\Phi$ is controlled by a $Z(3)$ symmetric effective potential ${\cal U}$, designed such that it reproduces the equation of state of pure gauge lattice QCD with its first order phase transition at a critical temperature of 270 MeV. The field $\phi$ acts as a potential on the quarks represented by the flavor doublet (for $N_f = 2$) or triplet (for $N_f = 3$) fermion field $\psi$. The Hamiltonian density in the quark sector is 
\begin{equation}
{\cal H} = -i\psi^\dagger(\vec{\alpha}\cdot\vec{\nabla} +\gamma_4\,\hat{m} - \phi)\psi + {\cal V}(\psi,\psi^\dagger)~,
\end{equation}
 with the quark mass matrix $\hat{m}$ and a chiral $SU(N_f)_L\times SU(N_f)_R$ symmetric interaction ${\cal V}$. 

Earlier two-flavor versions of the PNJL model \cite{Fu2003,RTW2006,RRW2007} have still used a local four-point interaction of the classic NJL type, requiring a momentum space cutoff to regularize loops. A more recent version \cite{HRCW2009}, using a non-local interaction between quarks, does not need any longer an artificial NJL cutoff. It generates instead a momentum dependent dynamical quark mass, $M(p)$, along with the non-vanishing quark condensate. A further extension to three quark flavors \cite{HRCW2010} includes a $U(1)_A$ breaking term implementing the axial anomaly of QCD. This term is constructed as a non-local generalization of the Kobayashi-Maskawa-'tHooft $3\times 3$ determinant interaction. 

A basic relation derived in this non-local PNJL model is the gap equation generating the momentum dependent quark mass self-consistently together with the chiral condensate $\langle\bar{\psi}\psi\rangle$. Its form (written here for the two-flavor case) is reminiscent of the corresponding equation emerging in Dyson-Schwinger approaches to QCD:
\begin{equation}
M(p) = m_0 + 8 N_c G\int{d^4 q\over (2\pi)^4} {\cal C}(p-q){M(q)\over q^2 + M^2(q)}~,
\end{equation}
where $m_0 = m_u = m_d$ is the current quark mass, $G$ is a coupling strength of dimension (length)$^2$ and ${\cal C}(q)$ is a momentum space distribution, the Fourier transform of which represents the range over which the
non-locality of the effective interaction between quarks extends in (Euclidean) 4-dimensional space-time. This non-local approach permits to establish contacts with the high-momentum limit of QCD with its well-known behaviour $M(p) \propto -\alpha_s(p)\langle\bar{\psi}\psi\rangle / p^2$ at $p\rightarrow \infty$. At 
$p \le 1$ GeV the distribution ${\cal C}(p)$ is designed to follow lattice QCD and Dyson-Schwinger results, 
or it may be deduced from an instanton liquid model (see Ref.~\cite{HRCW2010} for more details). The non-local PNJL model can indeed be ``derived" from QCD as demonstrated in Refs.~\cite{Ko2010,MP08}.

Once the input is fixed at zero temperature by well-known properties of the pseudoscalar mesons, the thermodynamics of the PNJL model can now be investigated with focus on the symmetry breaking pattern and on the intertwining of chiral dynamics with that of the Polyakov loop. The primary role of the Polyakov loop and its coupling to the quarks is to suppress the thermal distribution functions of color non-singlets, i.e. quarks and diquarks, as the transition temperature $T_c$ is approached from above. Color singlets, on the other hand, are left to survive below $T_c$. This is seen by analyzing the relevant piece of the thermodynamic potential $\Omega= - (T/V)\ln{\cal Z}$, the one involving the quark quasiparticles:
\begin{eqnarray}
{\Delta\Omega\over T} &\propto& \int {d^3p\over (2\pi)^3}\ln\left[1 + 3\Phi\, e^{-(E_p-\mu)/T} + 3\Phi^*\, e^{-2(E_p-\mu)/T} + e^{-3(E_p-\mu)/T}\right] \nonumber \\
& + & \int {d^3p\over (2\pi)^3}\ln\left[1 + 3\Phi^*\, e^{-(E_p+\mu)/T} + 3\Phi\, e^{-2(E_p+\mu)/T} + e^{-3(E_p+\mu)/T}\right] \,.
\end{eqnarray}
This suppression of color non-singlets in the hadronic phase below $T_c$ should, however, not be interpreted as dynamical confinement. The unsuppressed color singlet three-quark degrees of freedom are not clustered but spread homogeneously over all space. At this stage, nucleons are not treated properly as localized, strongly correlated three-quark compounds plus sea quarks. Thus one cannot expect that such a model describes properly the low-temperature phase of matter at finite baryon chemical potentials around 1 GeV, the domain of nuclear many-body systems.

%
\begin{figure}
\begin{center}
\includegraphics*[totalheight=8cm]{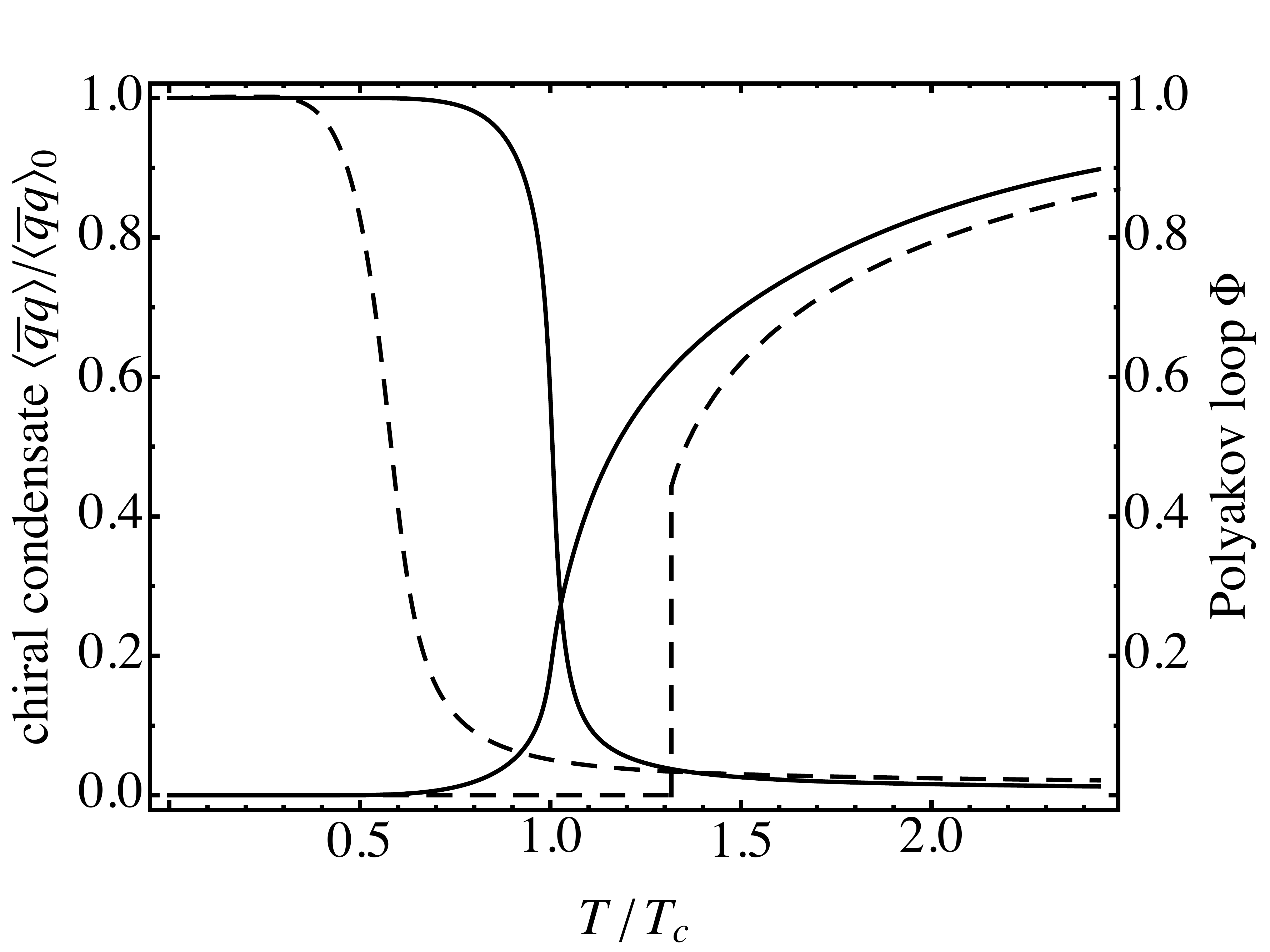}
\caption{Solid curves: temperature dependence of the chiral condensate $\langle\bar{q}q\rangle = \langle\bar{u}u
\rangle + \langle\bar{d}d\rangle$ (left) and of the Polyakov loop $\Phi$ (right) normalized to the transition temperature $T_c = 205$ MeV, as calculated in the nonlocal PNJL model with $N_f = 3$ and physical quark masses  \cite{HKW2011}. The dashed lines show the chiral condensate for the pure fermionic case and the Polyakov loop for the pure gluonic case, respectively.}
\label{Fig3}
\end{center}
\end{figure}
%

At zero baryon chemical potential, a remarkable dynamical entanglement of the chiral and deconfinement transitions is nonetheless verified in the PNJL model,  as demonstrated in Fig.\ref{Fig3}. In the absence of the Polyakov loop the quark condensate (left dashed line), shows the expected chiral crossover transition, but at a temperature way below and far separated from the 1st order deconfinement transition controlled by the pure-gauge Polyakov loop effective potential ${\cal U}$ (right dashed line). Once the coupling of the Polyakov loop field to the quark density is turned on, the two transitions move together and end up at a common transition temperature around 0.2 GeV. The deconfinement transition becomes a crossover (with $Z(3)$ symmetry explicitly broken by the coupling to the quarks).

It is interesting to compare this PNJL description of the interplay between the chiral and deconfinement transitions, with a scenario based on strong-coupling lattice QCD with inclusion of Polyakov loop dynamics \cite{NMO2010}. Although the starting points and frameworks are quite different in those two approaches, recent studies \cite{NMO2010} arrive at similar results concerning the close correspondence between chiral and deconfinement
transition temperature ranges. 

Pions also have a pronounced influence \cite{HRCW2010} on the chiral condensate as it approaches the transition region from the hadronic phase. The chiral crossover becomes significantly smoother close to $T_c$. Lattice QCD results \cite{Bo2010} with physical pion masses and realistic extrapolations to the continuum limit show a similar tendency.

Undoubtedly a prime challenge in the physics of strong interactions is the exploration of the QCD phase diagram at non-zero baryon density, extending from normal nuclear matter all the way up to very large quark chemical potentials $\mu_q$ at which color superconducting phases are expected to occur. PNJL calculations at finite $\mu_q$ give a pattern of the chiral order parameter in the $(T,\mu_q)$ plane showing a crossover at small $\mu_q$ that ends at a critical point. From there on a first-order transition line extends down to a quark chemical potential $\mu_q \sim$ 0.3 - 0.4 GeV at $T = 0$. 

\begin{figure}
\begin{center}
\includegraphics*[totalheight=8cm]{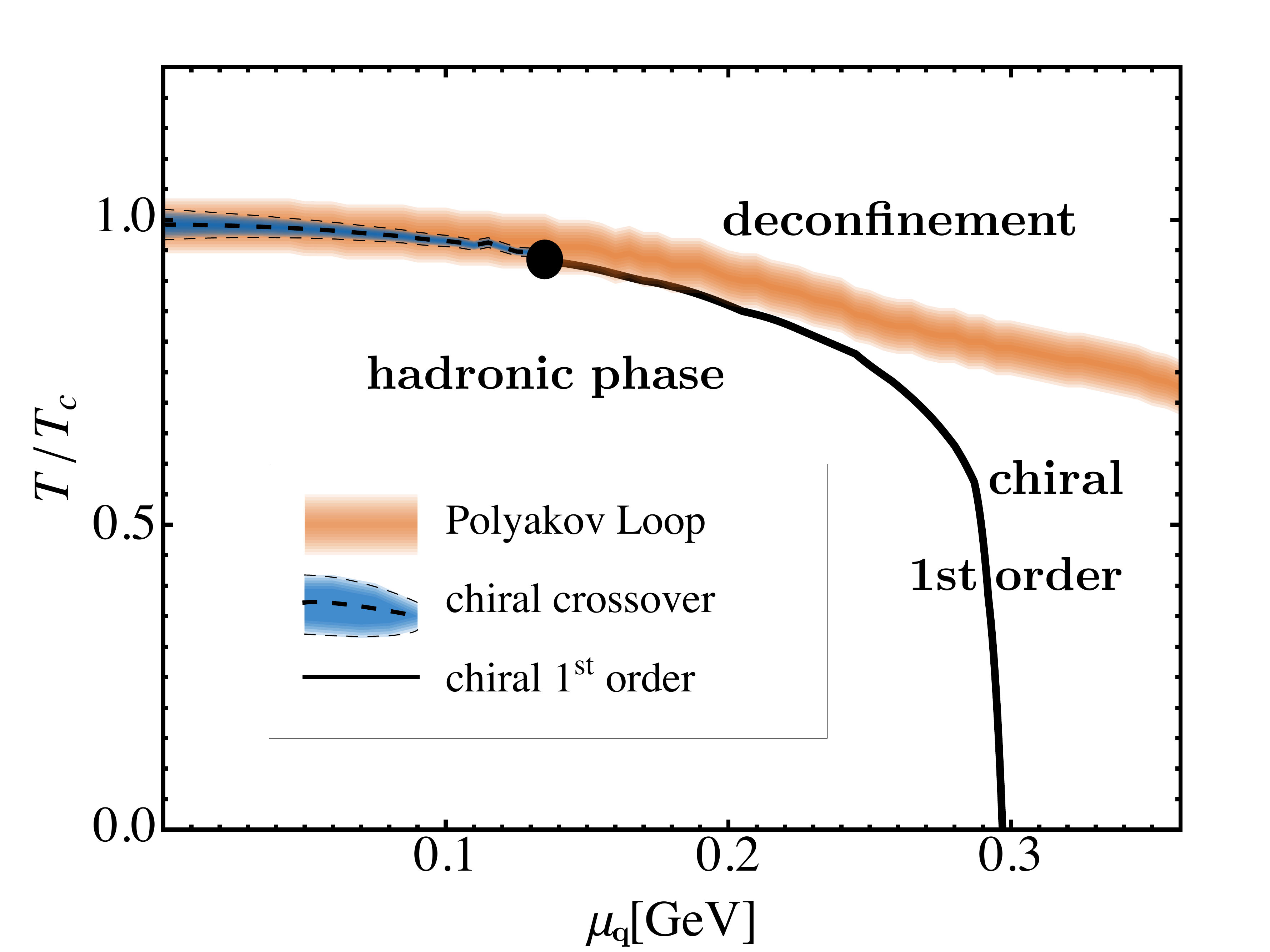}
\caption{Phase diagram for the three-flavor nonlocal PNJL model \cite{HKW2011}. The orange and blue bands
show the deconfinement and chiral crossover transitions. The solid black line indicates the chiral first-order transition. }
\label{Fig4}
\end{center}
\end{figure}

The typical phase diagram that emerges from the most recent version of the non-local three-flavor PNJL model \cite{HKW2011} is shown in Fig.\ref{Fig4}. The phase to the right of the 1st order transition line, but below the deconfinement boundary that decouples from the chiral transition beyond the critical point, has been named ``quarkyonic"\cite{HLP2008} and is hypothetically assumed to exist until superconducting phases take over at large quark chemical potential. A recent discussion about the possible relationship between such a hypothetical phase and the hadron rates produced in high-energy heavy-ion collisions \cite{An2010} may not yet be conclusive but is pursued with great intensity. 

%
%

One must recall again at this point that calculations and extrapolations based on PNJL or related approaches still have limited validity at moderate and high baryon densities since they do not properly incorporate baryonic degrees of freedom. They miss the important constraints imposed by the existence of nucleons and nuclear matter. This becomes clear when converting the PNJL phase diagram, Fig.\ref{Fig4}, into a plot that translates quark chemical potentials into baryon densities via the derivative of the pressure, $\rho_B = \partial P/\partial\mu_B$, with respect to baryon chemical potential $\mu_B = 3\mu_q$. The 1st order transition line converts to a broad coexistence region along the density axis that encloses what is well known as nuclear
physics terrain, but now described with the ``wrong" quasiparticle degrees of freedom: PNJL quarks rather than interacting nucleons. Quarks are not the relevant active quasiparticles at low temperature and baryon chemica potentials. Consequently, interpretations based on the schematic phase diagram of Fig.\ref{Fig4} along the chemical potential axis should be exercised with caution.  

Several further important questions are being raised in this context. A principal one concerns the existence and location of the critical point in the phase diagram. Extrapolations from lattice QCD, either by Taylor expansions\cite{Aoki2006} around $\mu =0$ or by analytic continuation from imaginary chemical potential \cite{FP2008}, have so far not reached a consistent conclusion. A second question relates to the sensitivity  of the first order transition line in the phase diagram with respect to the axial $U(1)_A$ anomaly in QCD \cite{Fu2008}. This issue has been addressed in Ref.~\cite{YTHB2007} using a general Ginzburg-Landau ansatz for the chiral SU(3) effective action with axial $U(1)$ anomaly. It was pointed out that, depending on details of the axial $U(1)_A$ breaking interaction, a second critical point might appear such that the low-temperature evolution to high density is again just a smooth crossover, or the first order transition might disappear altogether and give way to a smooth crossover throughout. In a three-flavor NJL type realization of this approach \cite{ABHY2010}, it has recently been demonstrated how the interaction between chiral condensate and diquark degrees of freedom, based on the genuine $U(1)_A$ breaking six-point vertex, can open up a smooth transition corridor along the  chemical potential axis in which chiral and diquark condensates may coexist.

%
\begin{figure}
\begin{center}
\includegraphics*[totalheight=8cm]{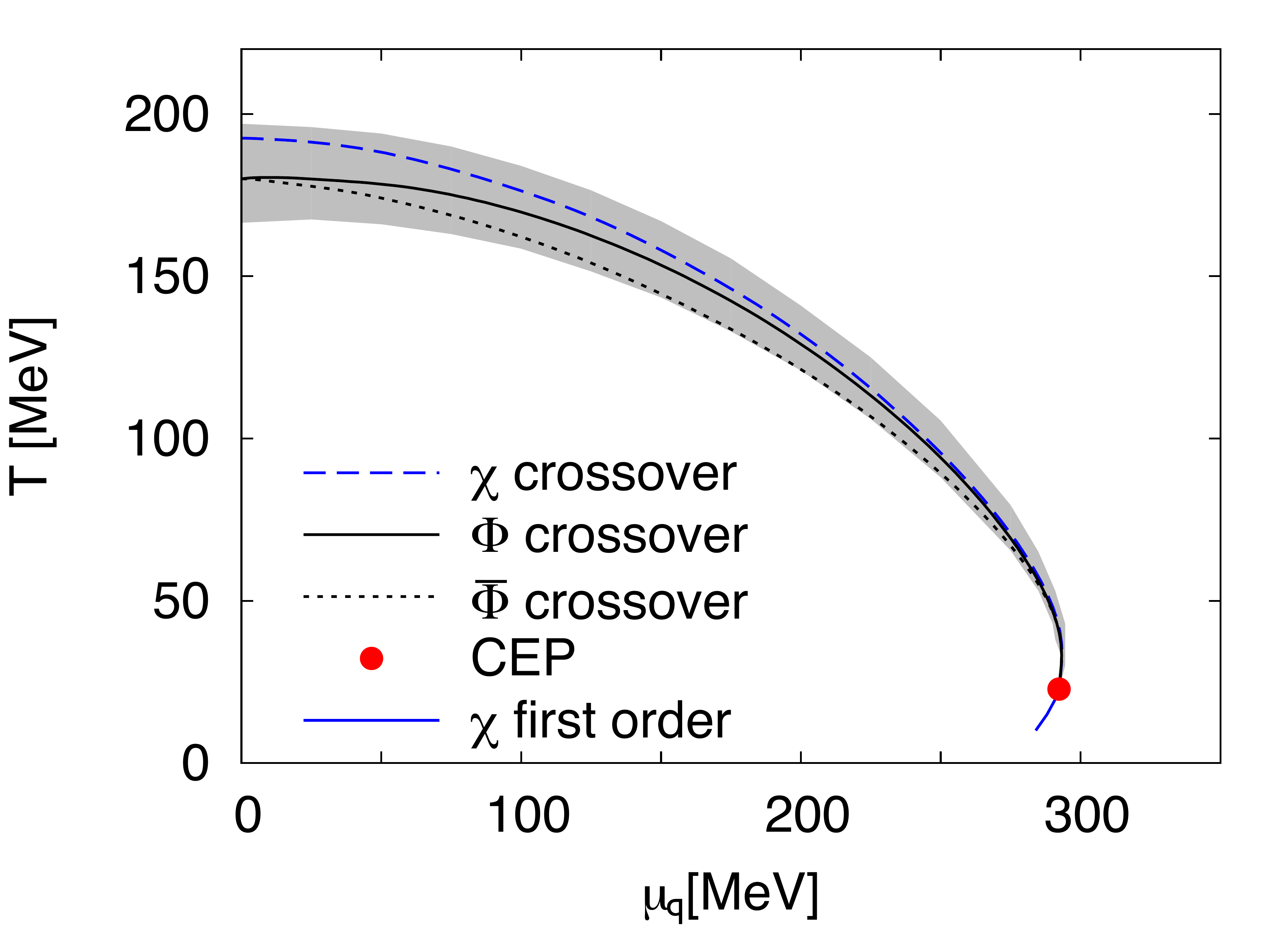}
\caption{Phase diagram for the two-flavor PQM (Polyakov - quark-meson) model \cite{HPS2011}. 
The solid and dotted lines show the deconfinement crossover transition ($\Phi$ and $\bar{\Phi}$ crossover),
while the dashed line marks the chiral crossover transition line ($\chi$ crossover) ending in the critical point.
The chiral first order transition is restricted to a small low-temperature region. 
 }
\label{Fig5}
\end{center}
\end{figure}
%

Interesting developments are presently conducted towards fixing the input of the non-local PNJL model by direct comparison with lattice QCD thermodynamics at imaginary chemical potential \cite{FP2008,Braun2011,KHW2011} and then studying implications on the phase diagram at real $\mu$.

\subsection{The Polyakov - quark-meson model}

A variant of the PNJL approach is the Polyakov - quark-meson (PQM) model \cite{PQM}. It is based on a linear sigma model with quarks coupled to chiral pion and sigma fields and propagating in a Polyakov loop background. Combined with renormalization group techniques, the PQM model has recently improved in a version \cite{HPS2011} that includes the back-reaction of the quarks at finite $\mu_q$ on the Polyakov loop effective potential. With increasing quark chemical potential, this back-reaction has the effect of aligning the chiral and deconfinement transition borders in the $T-\mu$ diagram such that these transitions tend to go in parallel not only at small chemical potential, but also as $\mu_q$ increases. This important feature is apparent in Fig.\ref{Fig5}. Note that the critical point is now moved to
low temperature and the chiral first order transition is restricted to a very small region around quark chemical potential $\mu_q \sim 0.3$ GeV in the phase diagram. However, once again, this region corresponds to the density range of nuclear matter where the quark-meson picture is not expected to be valid.

\subsection{Dyson-Schwinger approach to QCD thermodynamics}

\begin{figure}
\begin{center}
\includegraphics*[totalheight=8cm]{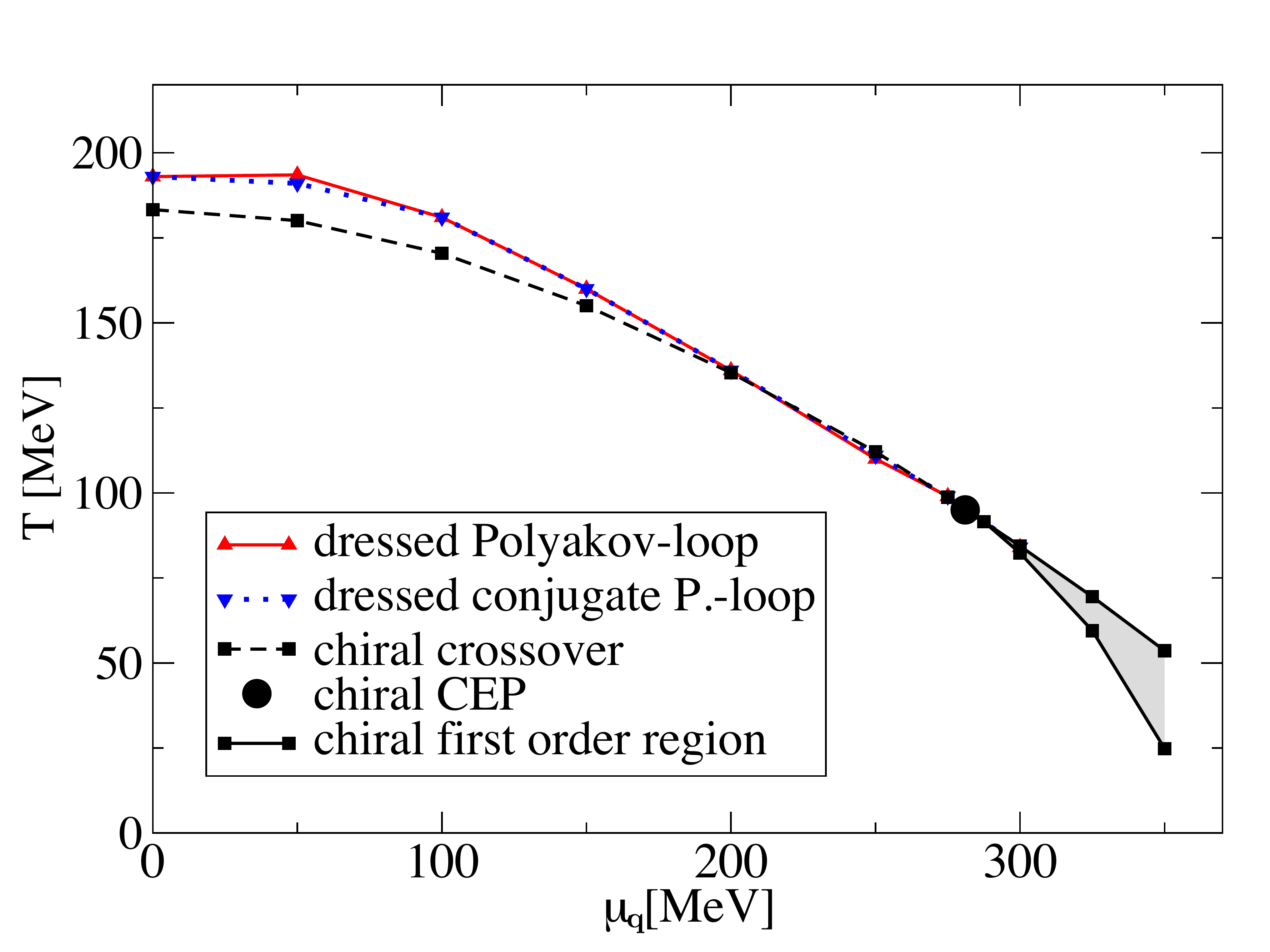}
\caption{Phase diagram derived from the Dyson-Schwinger approach to QCD thermodynamics \cite{FLM2011}. 
The curves show the chiral and deconfinement crossover transitions that turn out to almost coinicide
over the whole phase diagram. The chiral crossover ends at a critical point located at $T \simeq 100$ MeV and around $\mu_q \simeq 0.3$ GeV, below which the chiral transition becomes first order.
 }
\label{Fig6}
\end{center}
\end{figure}

The approach that is, by design and construction, closest to QCD itself is the one based on solutions of (truncated) Dyson-Schwinger equations for quark and gluon correlation functions. Dyson-Schwinger thermodynamics has now advanced to the point of producing its generic version of the QCD phase diagram \cite{FLM2011}, as displayed in Fig.\ref{Fig6}. It is again remarkable that the chiral and deconfinement crossovers go jointly in parallel from $T_c \simeq 170-180$ MeV at $\mu_q = 0$ down to a critical point at $T_e\simeq 100$ MeV and $\mu_q \simeq 280$ MeV. From there on downward a chiral first order transition region is indicated. 

%
%

What the Dyson-Schwinger and PQM approaches have in common, however, with the PNJL models is the absence of nucleon formation in the range of densities characteristic of nuclear (rather than constituent quark) matter.  The range of quark chemical potentials around $\mu_q \sim 0.3$ GeV and low temperatures is nuclear terrain! The baryon densities 
\begin{eqnarray}
\rho_B = {1\over 3} \left({\partial P\over \partial\mu_q}\right)_T~,
\end{eqnarray}
determined by the derivative of the pressure $P$ with respect to chemical potential, turn out to be in the range
$0.1 - 0.2$ fm, covering typical nuclear matter densities. Therefore models and approaches dealing with dynamical (constituent) quarks as quasiparticle degrees of freedom in that area cannot be correct: it is crucial to take into account the clustering of quarks into nucleons and their correlations in the nuclear medium.

\section{Nuclear chiral thermodynamics}

\begin{figure}
\begin{center}
\includegraphics*[totalheight=10cm]{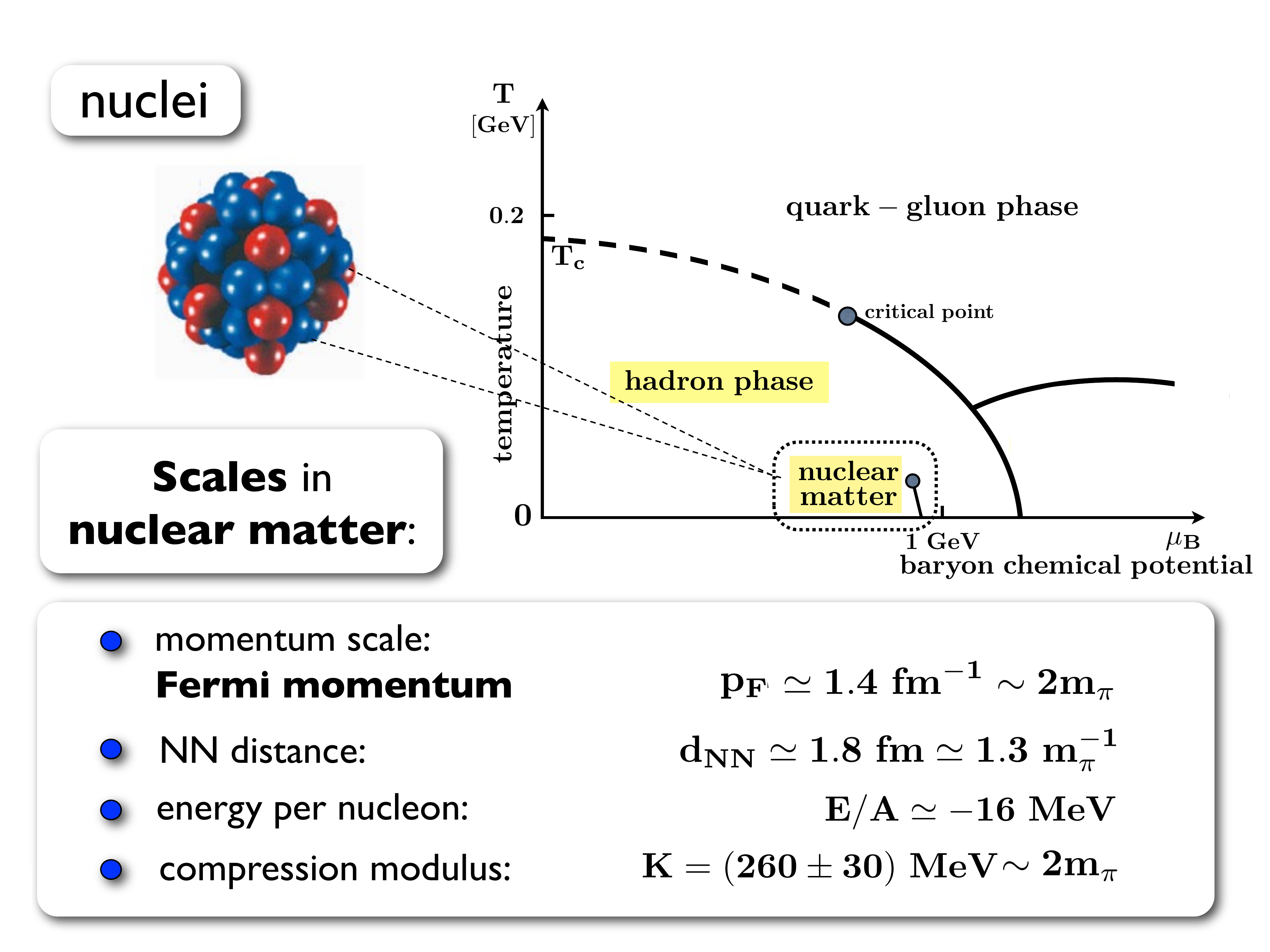}
\caption{Illustration of the position of nuclear matter in the QCD phase diagram, outlining 
relevant scales of the nuclear many-body problem.
 }
\label{Fig7}
\end{center}
\end{figure}

At this point it is now appropriate to turn towards the discussion of nuclear thermodynamics using the framework of chiral effective field theory, the low-energy realization of QCD in its sector with spontaneously broken chiral symmetry. As a starter, consider the characteristic scales at work in the ground state of symmetric nuclear matter, i.e. with equal number of neutrons and protons $(N = Z)$, illustrated in Fig.\ref{Fig7}. 

The nuclear many-body problem displays several ``small scales" compared to the spontaneous chiral symmetry breaking scale, $4\pi f_\pi \sim 1$ GeV. The Fermi momentum, $p_F \simeq 1.36$ fm$^{-1}$, is about half the inverse pion Compton wavelength, $\lambda_\pi = m_\pi^{-1}$. The corresponding average distance between two nucleons, $d_{NN} \simeq 1.8$ fm, is about 1.3 $\lambda_\pi$. The nuclear compression modulus, $\kappa = (260\pm 30)$ MeV, is about twice the pion mass. All these scales are ``pionic", i.e. of the same order as the pion Compton wave length or its inverse. The only exception is the binding energy per nucleon, $|E|/A \simeq 16$ MeV, an abnormally small number that reflects fine-tuning between short-range repulsion and intermediate-range attraction in the nucleon-nucleon force.

With these observations in mind, it is suggestive that the long- and intermediate-range behavior of the effective interactions responsible for properties of nuclear matter around its equilibrium density can be described by a chiral expansion (in-medium chiral perturbation theory) in powers of $p_F \ll 4\pi f_\pi \sim 1$ GeV. Short-distance correlations reflect large nucleon-nucleon scattering lengths and should instead be incorporated non-perturbatively by appropriate resummations or through a few contact interactions to be fitted to selected empirical data. 

Within the last decade such an approach to the nuclear many-body problem has been developed based on the understanding of the nucleon-nucleon interaction itself in terms of chiral effective field theory \cite{LFA2000,KFW2002,FKW2005}. In this approach, chiral one- and two-pion exchange processes in the nuclear medium are treated explicitly while unresolved short-distance dynamics are encoded in contact interactions. Three-body forces emerge naturally and play a significant role in this framework. The pion mass $m_\pi$, the nuclear Fermi momentum $p_F$ and the mass splitting $M_\Delta - M_N \simeq 2 m_\pi$  between the $\Delta(1232)$ and the nucleon are all comparable ``small" scales that figure as expansion parameters. The relevant,
active degrees of freedom at low energy are therefore pions, nucleons and $\Delta$ isobars. Two-pion exchange interactions produce intermediate-range Van der Waals - like forces involving the large spin-isospin polarizablity of the individual nucleons. The Pauli principle plays an important role acting on intermediate nucleons as they propagate in two-pion exchange processes within the nuclear medium. 

%
%

\subsection{In-medium chiral perturbation theory}

A key ingredient of chiral perturbation theory applied to nuclear matter is the in-medium nucleon propagator,
\begin{equation}
S_N(p) = (\gamma_\mu p^\mu + M_N) \left[{i\over p^2 - M_N^2 + i\varepsilon} - 2\pi\,\delta(p^2 - M_N^2)
\,\theta(p^0)\,\theta(p_F -|\vec{p}\,|)\right]~,
\label{Eq:Nprop}
\end{equation}
that takes into account effects of the filled nuclear Fermi sea (with Fermi momentum $p_F$). These propagators enter in the loop diagrams generating the free energy density. Thermodynamics is introduced using the Matsubara formalism. The free energy density is then computed as a function of temperature $T$ and baryon density $\rho$ at given Fermi momentum $p_F$. It is derived in the form reminiscent of a virial expansion:
\begin{eqnarray} 
{\cal F}(\rho,T)= 4\int_0^\infty dp\,p \, {\cal K}_1(p)\,D(p)
&+&\int_0^\infty dp_1\int_0^\infty dp_2\,{\cal K}_2(p_1, p_2)\, D(p_1)
D(p_2)\nonumber \\ 
&+& \int_0^\infty dp_1\int_0^\infty dp_2\int_0^\infty dp_3\, {\cal K}_3(p_1, p_2, p_3) \,D(p_1) D(p_2)D(p_3) + . . . 
\label{FreeEnergy}
\end{eqnarray}
Complete expressions for the kernels ${\cal K}_i$ are given in Refs.\cite{FKW2005,FKW2011}. For example, the one-body kernel ${\cal K}_1$ represents the contribution of the non-interacting nucleon gas, including relativistic corrections, to the free energy density and it reads 
\begin{equation} \label{K1}
{\cal K}_1 = M_N +\tilde \mu- {p_1^2\over 3M_N}- {p_1^4\over 8M_N^3} \,. 
\end{equation}
The terms involving ${\cal K}_2$ and ${\cal K}_3$ describe the effects of interactions and include two- and three-nucleon correlations, and so forth.
The quantity  
\begin{equation} 
D(p) = {p\over 2\pi^2} \bigg[ 1+\exp{p^2/2M_N -\tilde \mu \over T} 
\bigg]^{-1} \,,
\end{equation}
%
\begin{figure}[htb]
\begin{minipage}[t]{9cm}
\includegraphics[width=8cm]{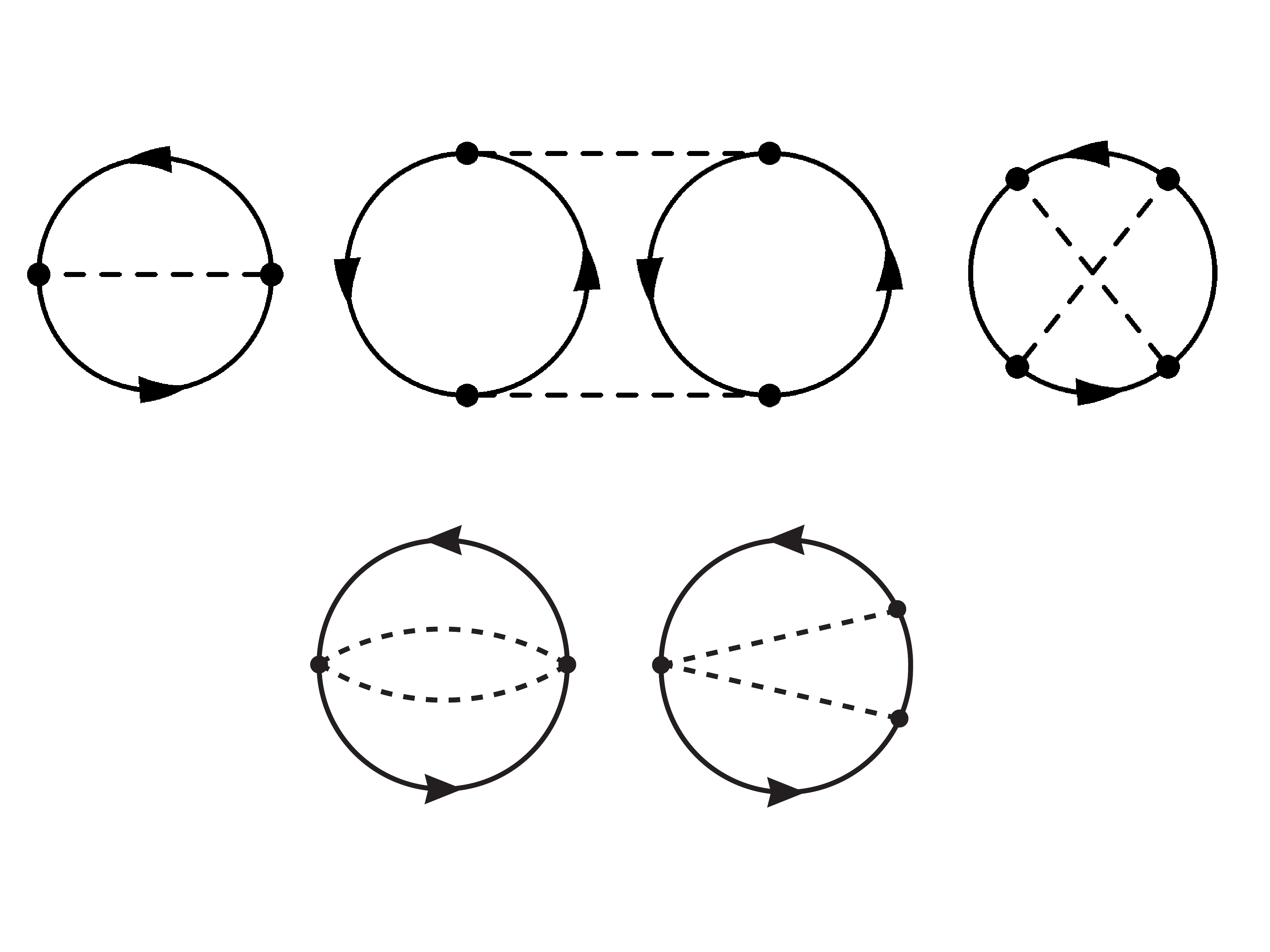}
\caption{One- and two-pion exchange processes contributing to the ground state energy (or the free energy density at non-zero temperature). Each nucleon line is subject to a medium insertion (see Eq.(\ref{Eq:Nprop})). } 
\label{Fig8}
\end{minipage}
\hspace{\fill}
\begin{minipage}[t]{9cm}
\includegraphics[width=8cm]{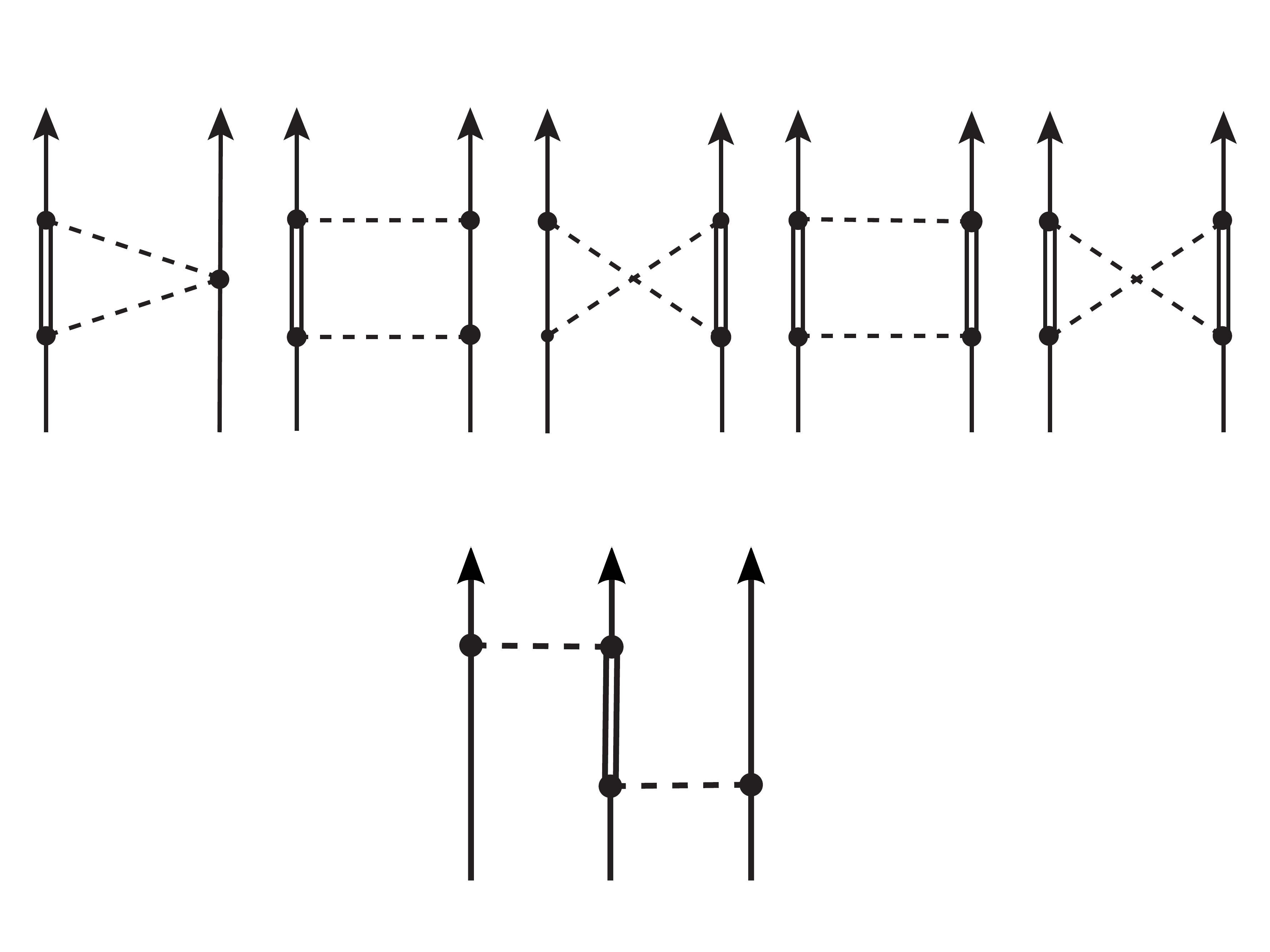}
\caption{Two-pion exchange processes involving one or two $\Delta(1230)$ isobars in intermediate states. The lower part of the figure shows the corresponding 3-body interaction induced by virtual $\Delta(1230)$ excitation.} 
\label{Fig9}
\end{minipage}
\end{figure}
%
denotes the density of nucleon states in momentum space, written as a product of 
the temperature dependent Fermi-Dirac distribution and a kinematical prefactor 
$p/ 2\pi^2$ that has been included in $D(p)$  for convenience. 
$M_N$ stands for the (free) nucleon mass. The baryon density $\rho$ is 
calculated as
\begin{equation}
\rho= 4\int_0^\infty  dp\,p\,D(p)\,. 
\end{equation} 
This relationship determines the dependence of the effective one-body 
``chemical potential" $\tilde \mu(\rho,T;M_N)$ on the thermodynamical variables 
$(\rho, T)$. The ``true" nucleon chemical potential that figures as a variable in the phase diagram, is determined by 
$\mu = \bar{F} + \rho(\partial\bar{F}/\partial\rho)$ where $\bar{F} = {\cal F}/\rho$ is the free energy per nucleon. The pressure is calculated using the standard relation
\begin{equation}
P(\rho,T) = \rho^2\,{\partial\bar{F}(\rho,T) \over \partial\rho}~.
\end{equation} 

The loop expansion of in-medium chiral perturbation theory is related in a one-to-one correspondence to a systematic expansion of the energy density in powers of $p_F$, with expansion coefficients expressed as functions of the dimensionless ratio $m_\pi/p_F$.  
Calculations of nuclear thermodynamics have been performed up to and including three-loop order in the energy density. A typical set of
pion-exchange diagrams is displayed in Fig.\ref{Fig8}. Also explicitly included in the calculation are intermediate virtual $\Delta(1232)$ excitations and resulting three-body terms shown in Fig.\ref{Fig9}.  The three-loop truncation in the chiral expansion of the pressure in powers of the Fermi momentum, $p_F$, valid unless four-nucleon correlations are substantial, implies that these calculations can be safely trusted up to about twice the density of normal nuclear matter, i.e. for $p_F$ less than $0.3$ GeV $\ll 4\pi f_\pi$, with $f_\pi = 0.09$ GeV the pion decay constant in vacuum. 

\begin{figure}
\begin{center}
\includegraphics*[totalheight=7cm]{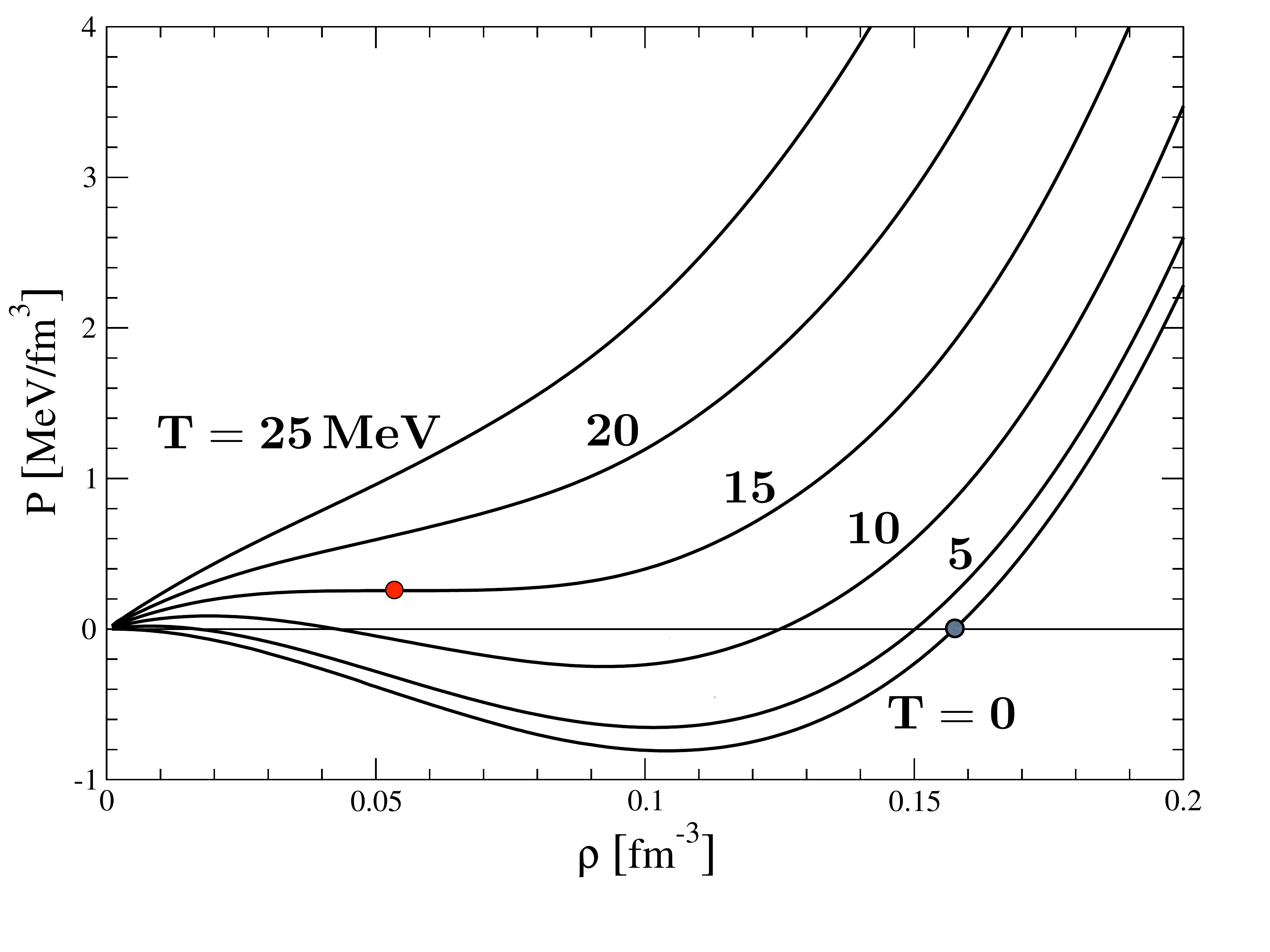}
\caption{Equation of state of symmetric $(N=Z)$ nuclear matter from in-medium chiral effective field theory \cite{FKW2005,FKW2011}. The critical point for the first-order liquid-gas phase transition appears at $T_c \simeq 15$ MeV. The equlibrium point at $T = 0$ and baryon density $\rho_0 \simeq 0.16$ fm$^{-3}$ is fixed by a single parameter, the strength of an NN contact interaction. All remaining input is pre-determined by known pion-nucleon interactions in vacuum.
 }
\label{Fig10}
\end{center}
\end{figure}

\begin{figure}[htb]
\begin{minipage}[t]{9cm}
\includegraphics[width=8cm]{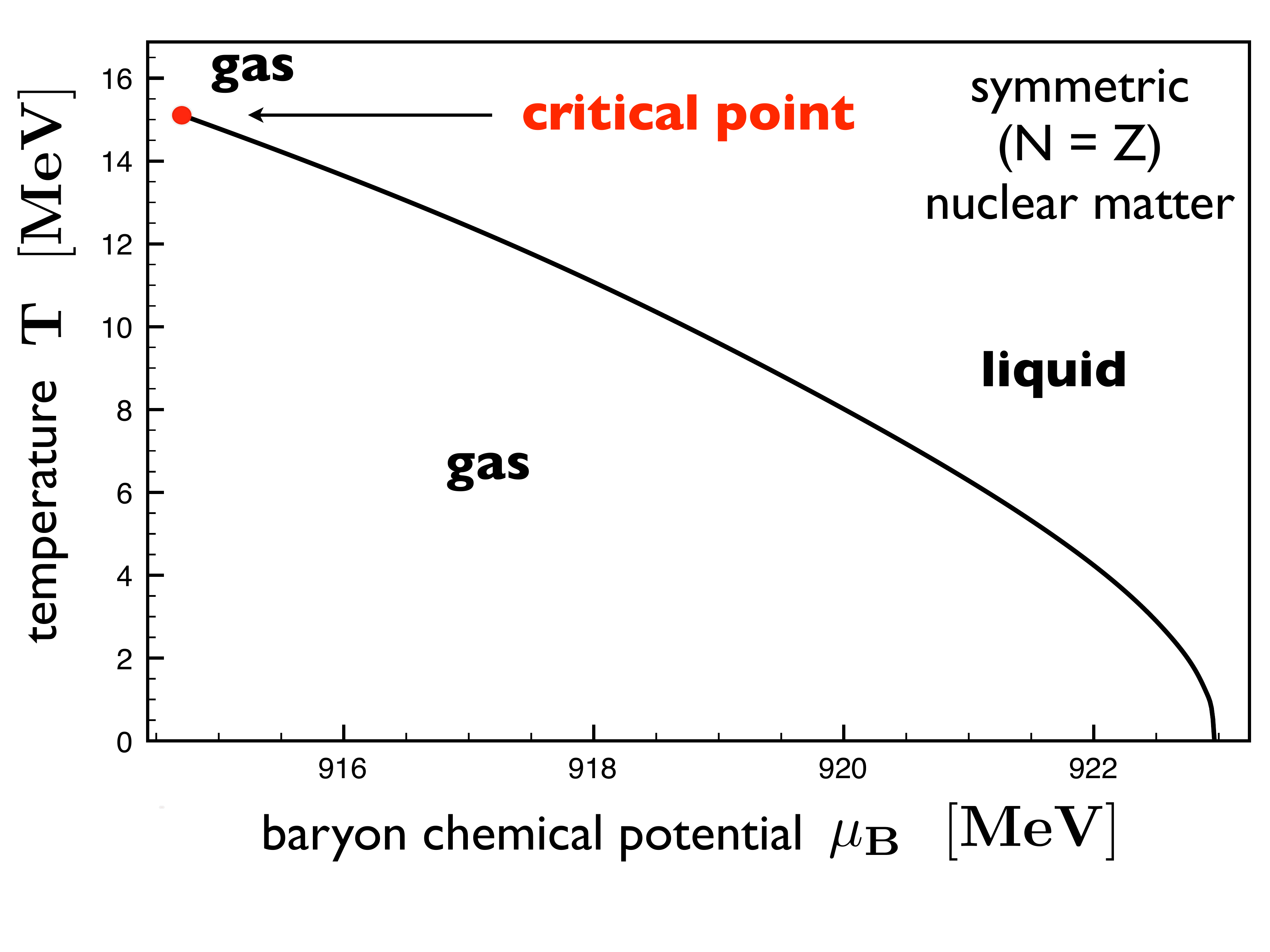}
\caption{Phase diagram of symmetric $(N=Z)$ nuclear matter from in-medium chiral effective field theory \cite{FKW2011}.} 
\label{Fig11}
\end{minipage}
\hspace{\fill}
\begin{minipage}[t]{9cm}
\includegraphics[width=8cm]{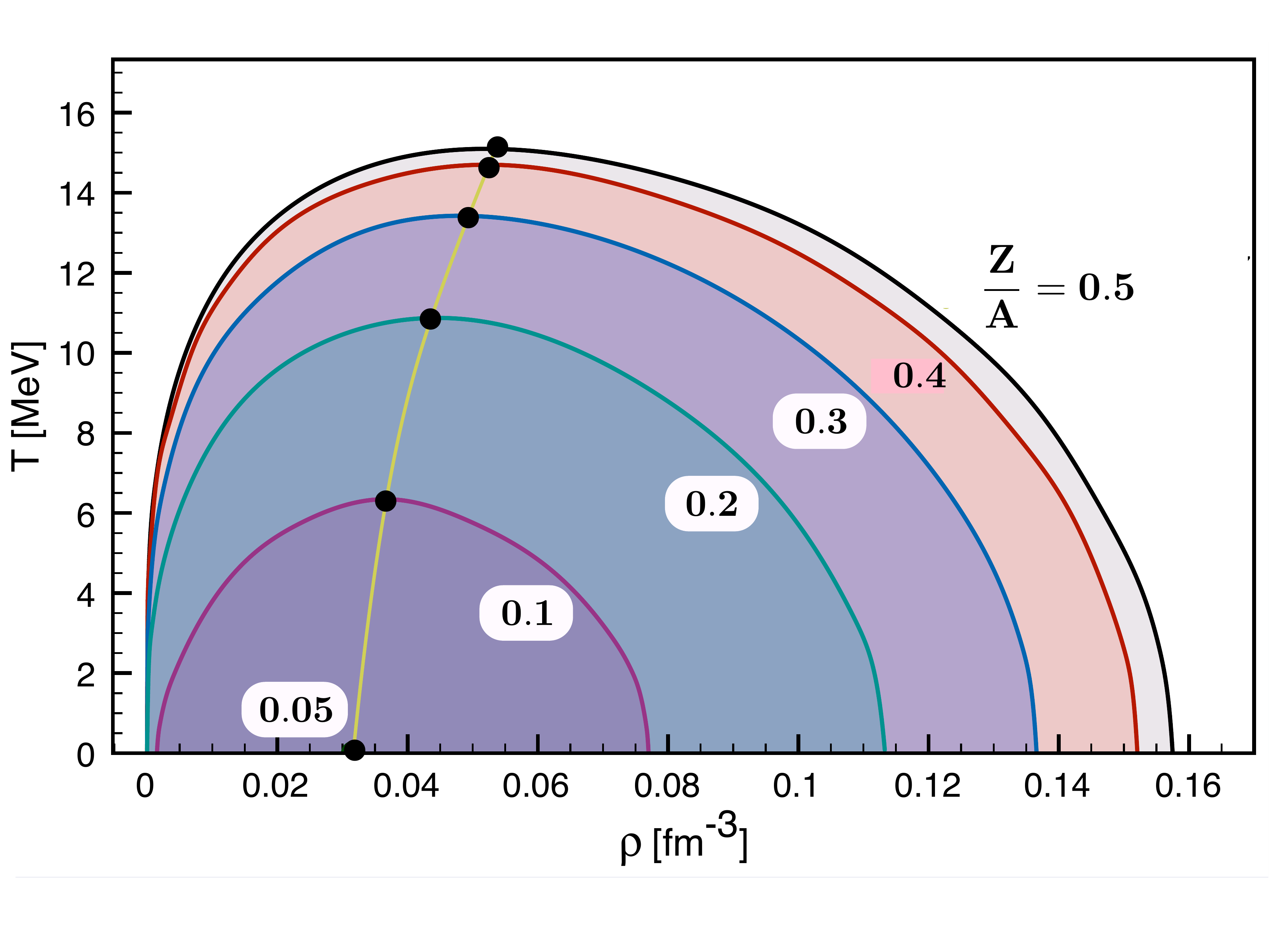}
\caption{Evolution of the liquid-gas coexistence region from symmetric nuclear matter ($Z/A = 0.5$) to neutron-rich matter calculated using in-medium chiral effective field theory \cite{FKW2011}.} 
\label{Fig12}
\end{minipage}
\end{figure}

A limited set of constants associated with $NN$ contact terms is fixed by reproducing e.g. the binding energy per nucleon in equilibrium nuclear matter at $T=0$. Then this scenario leads to a realistic nuclear matter equation of state \cite{FKW2005} (see Fig.\ref{Fig10}) with a liquid-gas first order phase transition and a critical temperature of about 15 MeV, close to the range of empirical values extracted for this quantity. Incidentally, this nuclear liquid-gas transition is so far the only well established part of the phase diagram for strongly interacting matter at finite density. 
%
%
Fig.\ref{Fig11} shows the ($T,\mu_B$) phase diagram for $N=Z$ nuclear matter, while Fig.\ref{Fig12} displays ($T,\rho$) phase diagrams for isospin-asymmetric nuclear matter, outlining the evolution of the critical point and the liquid-gas coexistence region for increasing proton fractions $Z/A$ \cite{FKW2011}. The first order phase transition stops existing at $Z/A=0.05$, at which point the liquid component disappears and neutron matter begins to be realized as an interacting Fermi gas. It is important to note that this behaviour is almost entirely driven by the isospin dependence of the in-medium two-pion exchange interactions. Typical ingredients of one-boson exchange phenomenology (such as rho and sigma bosons) do not appear in this approach and are replaced by explicit $2\pi$ exchange mechanisms.
%

Extensions of this framework to finite nuclei, using the energy density functional formalism, have been
applied successfully \cite{FKVW2006} to calculate ground state properties (binding energies, mean square radii, deformations and systematics along isotopic chains) throughout the nuclear chart,  from $^{16}O$ to $Pb$ isotopes. Other examples of detailed nuclear structure phenomena such as the anomalously weak $^{14}C$ beta decay transition, with its lifetime of almost six thousand years that enables radiocarbon dating, can be understood in terms of the density dependent chiral effective interaction including a pronounced three-body contribution \cite{HKW2009}.   
Further recent applications include the investigation of chiral NN interactions in the context of nuclear energy density functionals and the quasiparticle interaction in Fermi liquid theory \cite{HKW2011a}.

\subsection{Density and temperature dependence of the chiral condensate}

In a nuclear equation of state based on chiral dynamics the pion mass enters explicitly (or, equivalently, the quark mass $m_q$ according to the Gell-Mann - Oakes - Renner relation,  $m_\pi^2 f_\pi^2 = -m_q\langle\bar{\psi}\psi\rangle$). Equiped with such an equation of state one can now ask the following questions: how does the chiral (quark) condensate $\langle\bar{\psi}\psi\rangle$ extrapolate
to baryon densities $\rho$ exceeding those of normal nuclear matter, and how does it evolve as a function of temperature $T$ in a nuclear environment? The answer, first at $T=0$, is based on the Hellmann-Feynman theorem applied to the Hamiltonian density of QCD in the nuclear ground state $|\Psi\rangle$, 
\begin{equation}
\langle\Psi|\bar{\psi}\psi|\Psi\rangle = \langle\Psi|{\partial{\cal H}_{QCD}\over\partial  m_q}|\Psi\rangle = \langle\bar{\psi}\psi\rangle_0\left[1 - {\partial{\cal E}(m_\pi;\rho)\over f_\pi^2\,\partial m_\pi^2}\right]~~.
\label{eq:condensate}
\end{equation}
where $\langle\bar{\psi}\psi\rangle_0$ is the chiral condensate in vacuum.  The energy density ${\cal E}$ (normalized so that it vanishes at $\rho \rightarrow 0$) is expressed at given baryon density $\rho$ as a function of the
pion mass. In-medium chiral effective field theory gives the following result \cite{KHW2008}:
\begin{eqnarray}
{\langle\bar{\psi}\psi\rangle_\rho\over\langle\bar{\psi}\psi\rangle_0} = 1&-&{\rho\over f_\pi^2}{\sigma_N\over m_\pi^2}\left(1 - {3\,p_F^2\over 10\,M_N^2} + \dots\right) - {\rho\over f_\pi^2}{\partial\over \partial m_\pi^2}\left({E_{int}(m_\pi; p_F)\over A}\right)~. 
\end{eqnarray}
The second term on the r.h.s., with its leading linear dependence on density, is the contribution from a free Fermi gas of nucleons. It involves the pion-nucleon sigma term $\sigma_N\simeq 0.05$ GeV and additional non-static corrections. The third term includes the pion mass dependence of the interaction energy per nucleon, $E_{int}/A$. This term features prominently the two-pion exchange interaction in the nuclear medium  including Pauli principle corrections, and also three-nucleon forces based on two-pion exchange.
%
\begin{figure}
\begin{center}
\includegraphics*[totalheight=7.5cm]{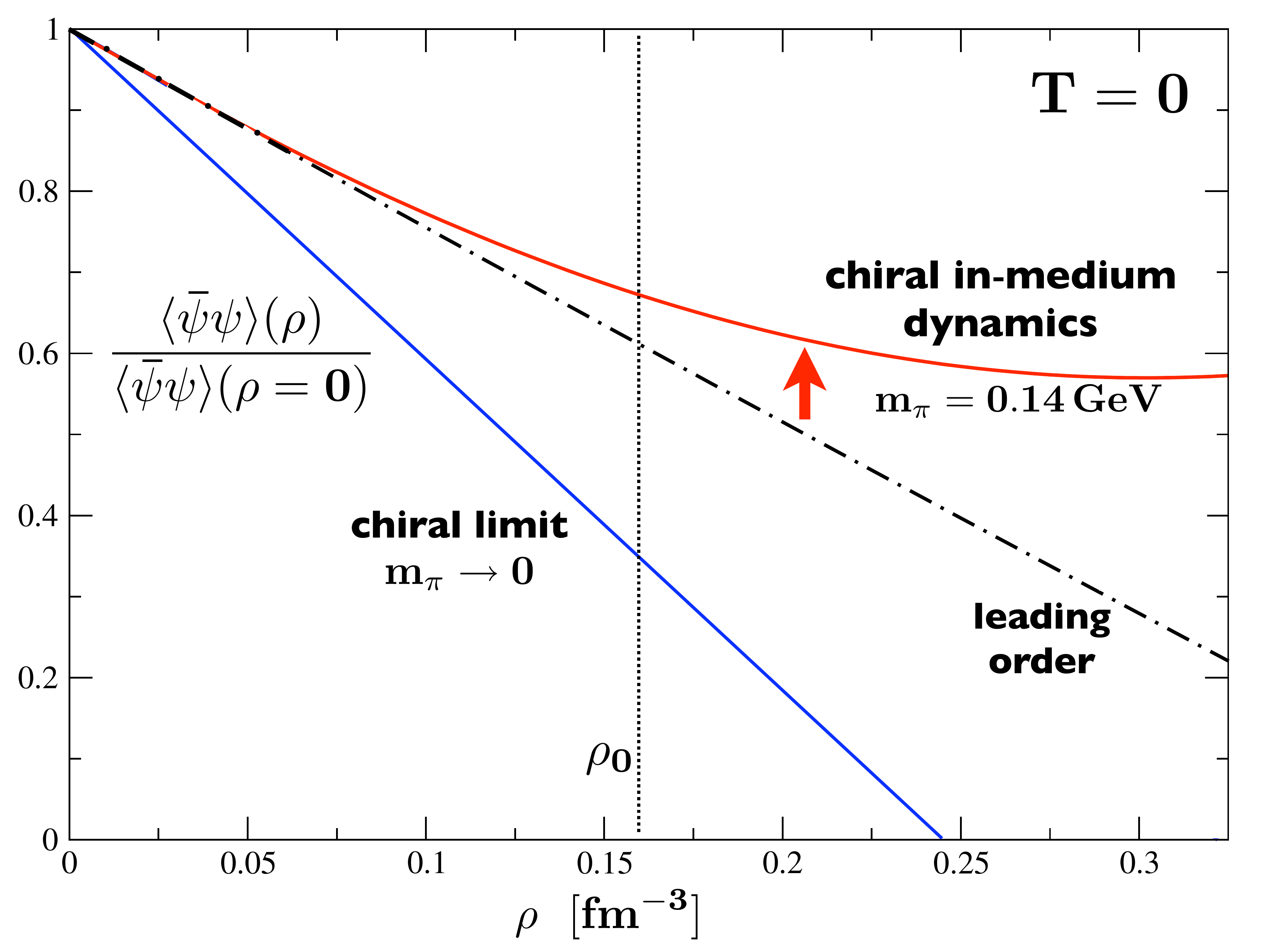}
\caption{Density dependence of the chiral condensate at temperature $T = 0$ in symmetric nuclear matter \cite{KHW2008}. Dot-dashed curve: leading order term using $\sigma_N = 50$ MeV. Upper curve: full in-medium chiral effective field theory  result at three-loop order. Lower curve: chiral limit with vanishing pion mass.
 }
\label{Fig13}
\end{center}
\end{figure}
%
The dashed-dotted curve in Fig.\ref{Fig13} shows the pronounced leading linear reduction in the magnitude of the chiral condensate with increasing density. This holds in the absence of correlations between the nucleons. Up to about the density of normal nuclear matter, this term dominates, whereas the interaction part tends to delay the tendency towards chiral restoration when the baryon density is further increased. 
This stabilizing effect arises from the combination of Pauli blocking in two-pion exchange processes and
three-body correlations, and their explicit dependence on the pion mass. This behaviour is highly sensitive to the actual value of the pion mass. In the chiral limit, $m_\pi\rightarrow 0$, with stronger attraction in the NN force at intermediate ranges, the trend is reversed and the more rapidly dropping condensate would now lead to the restoration of chiral symmetry at relatively low density. In essence, nuclear physics would look radically different if pions were exactly massless Nambu-Goldstone bosons.  The drastic influence of explicit chiral symmetry breaking in QCD through the small but non-zero $u$ and $d$ quark masses on qualitative properties of nuclear matter is quite remarkable. 

%
\begin{figure}[htb]
\begin{minipage}[t]{8.5cm}
\includegraphics[width=8.5cm]{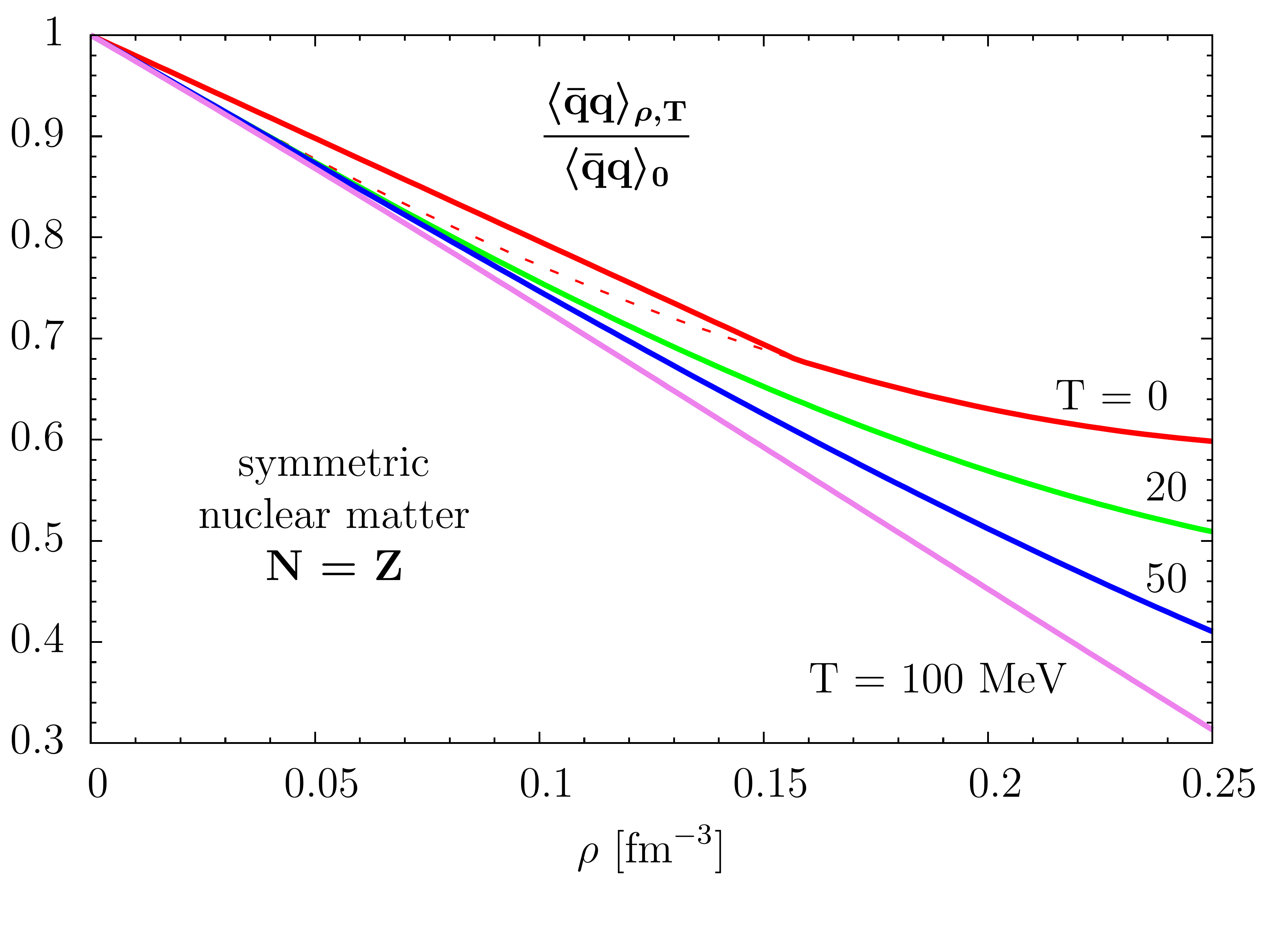}
\caption{Density and temperature dependence the chiral condensate in symmetric nuclear matter \cite{FKW2011}. At $T = 0$ the influence of the first-order liquid-gas phase transition is indicated by drawing a straight line corresponding the Maxwell construction (ignored in Fig.\ref{Fig13}).} 
\label{Fig14}
\end{minipage}
\hspace{\fill}
\begin{minipage}[t]{8.5cm}
\includegraphics[width=8.5cm]{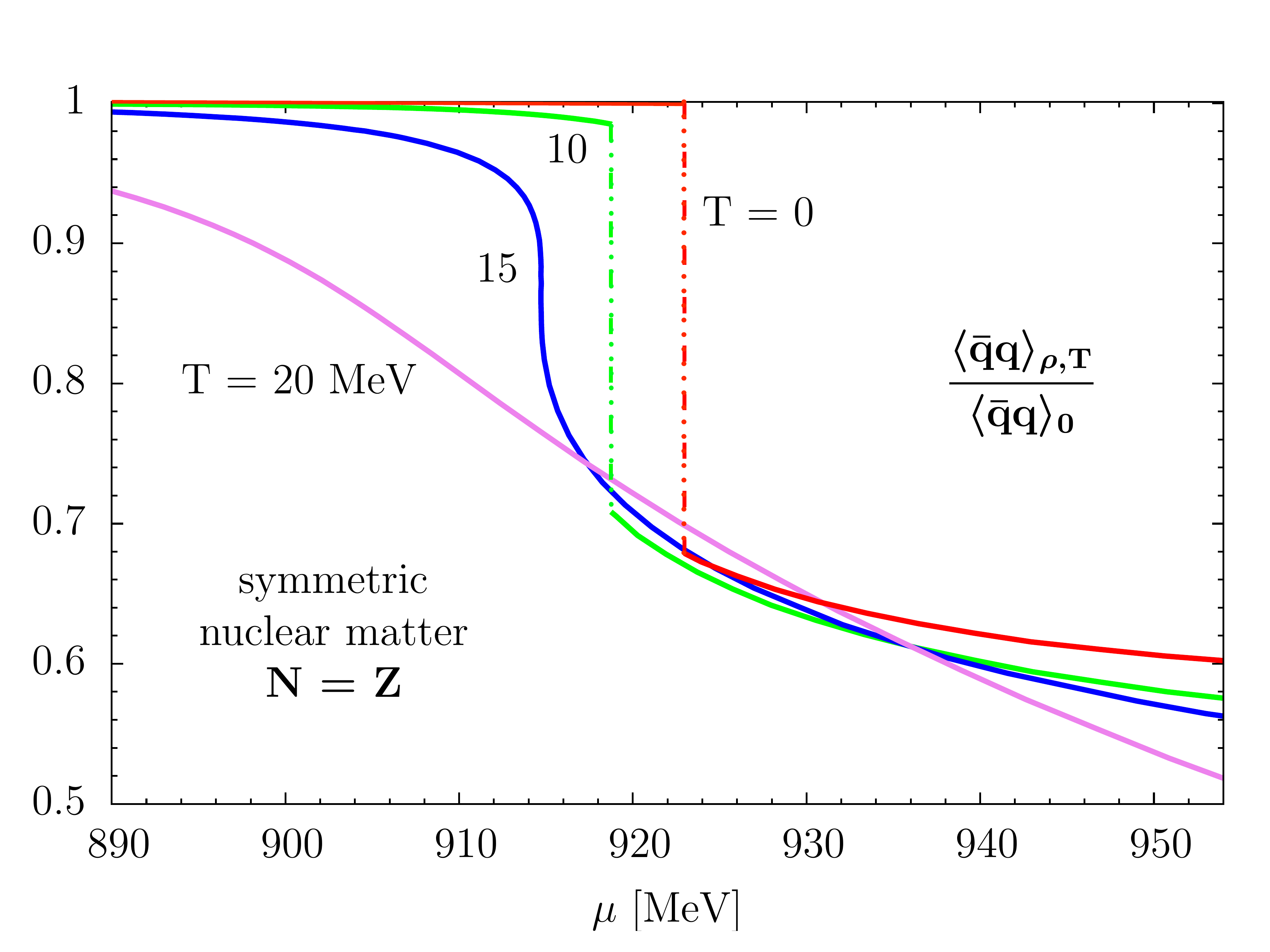}
\caption{Temperature dependence of the chiral condensate  \cite{FKW2011} in symmetric nuclear matter as function of the baryon chemical potential $\mu$. This plot demonstrates the pronounced influence of the first-order liquid-gas phase transition at low temperatures.} 
\label{Fig15}
\end{minipage}
\end{figure}
%

The additional temperature dependence of the in-medium chiral condensate \cite{FKW2011a} is found by repeating the calculation, Eq.(\ref{eq:condensate}), using the free energy density, ${\cal F}(m_\pi; \rho, T)$:
\begin{equation}
{\langle\bar{\psi}\psi\rangle_{\rho, T}\over \langle\bar{\psi}\psi\rangle_0} = 1 -  {\partial{\cal F}(m_\pi;\rho, T))\over f_\pi^2\,\partial m_\pi^2}~~.
\end{equation} 
The result shown in Fig.\ref{Fig14} demonstrates how the chiral condensate ``melts" with increasing temperature and approaches the linear density dependence of the non-interacting Fermi gas at a temperature of about 100 MeV.
At low temperatures, the influence of the nuclear liquid-gas first-order phase transition is visible even in the chiral condensate. This becomes particularly pronounced when the condensate is plotted as a function of baryon chemical potential $\mu$ (see Fig.\ref{Fig15}). The effect of thermal pions on the temperature dependence of the quark condensate must still be added; it amounts \cite{GL89, Kai99} to lowering the magnitude of the condensate by another 5\% of its vacuum value at $T\simeq 100$ MeV. 

However, no tendency towards a first order chiral phase transition is visible, at least up to baryon densities about twice that of normal nuclear matter and up to temperatures $T \sim 100$ MeV. In that range it appears that the relevant fermionic quasiparticle degrees of freedom are indeed nucleons rather than quarks. This should be kept in mind as an important constraint for the discussion of the QCD phase diagram at finite baryon chemical potential.  

\section{New constraints from neutron stars}

The equation-of-state of highly compressed cold matter in the center of neutron stars has been an issue and a challenge for many decades. For a long time it was thought that the vast majority of neutron stars have masses
around $1.5$ times the solar mass. The equation-of-state of neutron star matter that serves as an input for the Tolman-Oppenheimer-Volkov equation to calculate the mass-radius relation of the star, $M(R)$, should then be sufficiently ``stiff" in order to support such neutron star masses. Many scenarios for building the cores of such compact stars were designed and discussed \cite{LP2007}, from conventional nucleonic matter equations-of-state \cite{APR1998} to matter with more exotic components such as kaon condensates and quark matter \cite{Sch2010}.   

The empirical constraints were considerably sharpened recently when a neutron star in a binary with a white dwarf was observed. An extremely accurate Shapiro delay measurement of the pulsar signal resulted in the unusually large mass of $1.97\pm 0.04$ in solar mass units \cite{Dem2010}. This observation rules out some of the softer equations-of-state, e.g. those featuring kaon condensation.
%
\begin{figure}
\begin{center}
\includegraphics*[totalheight=7.8cm]{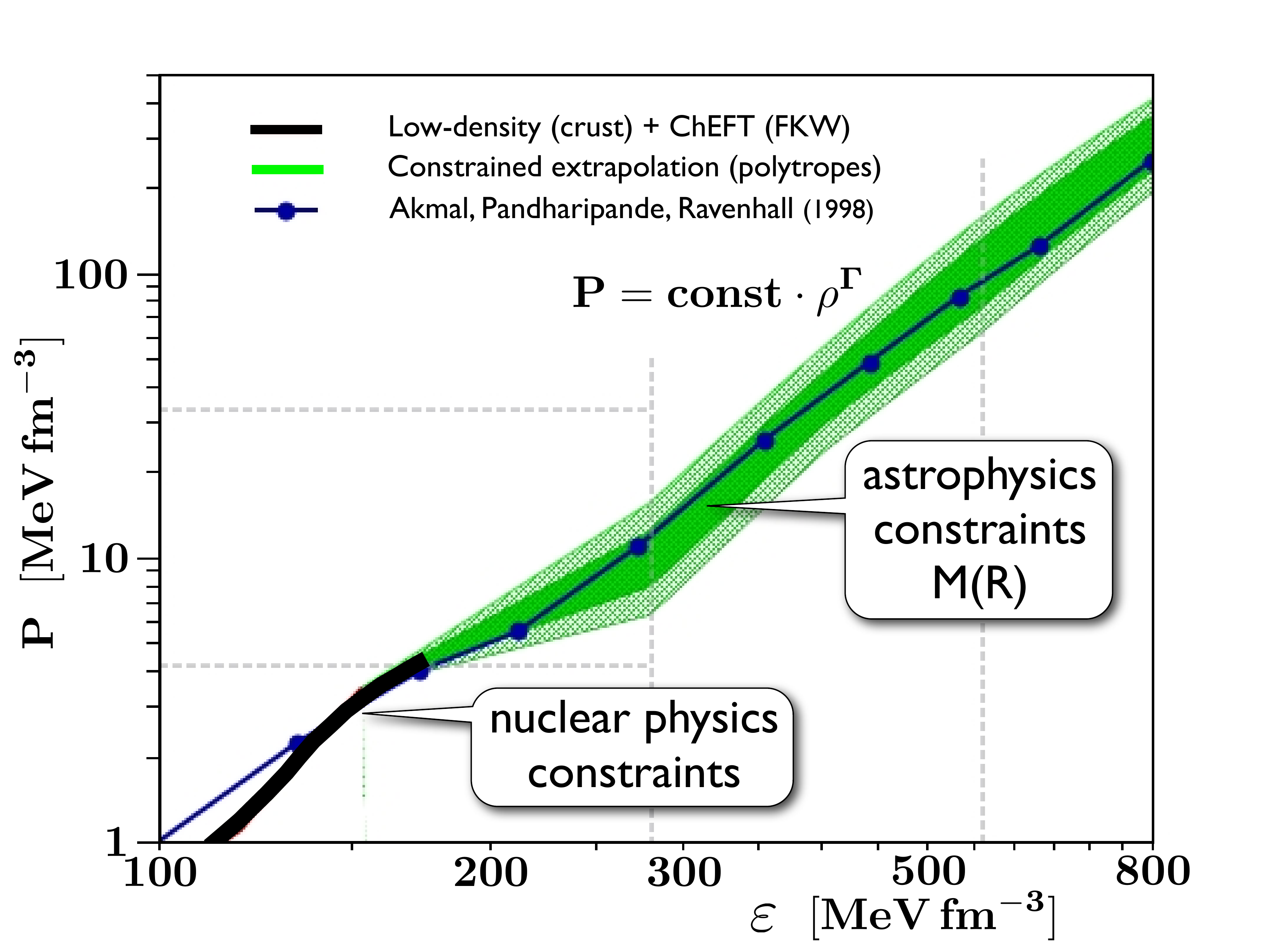}
\caption{Pressure versus energy density for neutron star matter (uncertainty band included) \cite{RW2011} with built-in constraints resulting from the existence of a two-solar-mass neutron star \cite{Dem2010} and from the analysis of most probable mass-radius relations, $M(R)$, from Ref. \cite{SLB2010}. The black solid area is the neutron matter equation of state (with 10\% protons) calculated using in-medium chiral effective field theory to three-loop order \cite{FKW2011}. The solid line is the APR equation of state \cite{APR1998} based on a realistic nuclear many-body calculation.
 }
\label{Fig16}
\end{center}
\end{figure}
%

A further important constraint comes from the systematic analysis of a well examined series of neutron stars. Using this data base, a most probable mass-radius relation $M(R)$ has been established \cite{SLB2010}, with a relatively narrow window of star radii around $R \simeq 12$ km. The combination of all these pieces of information
can now be used in order to set rather tight constraints for the equation-of-state of dense neutron star matter. One starts out at low densities with a reliable equation-of-state for the crust of the star and then proceeds to higher densities
using a realistic neutron matter equation-of-state based on the chiral effective field theory calculations discussed earlier \cite{FKW2005,FKW2011,HLPS2010}. Such an extrapolation holds up to baryon densities of about twice that of normal nuclear matter. The further extrapolation to high densities is then performed using sections of polytropes, parametrizing the pressure as $P = const.\,\rho^{\Gamma}$. When this is all done one arrives at an equation-of-state,
i.e. pressure $P = P(\cal{E})$ as a function of energy density $\cal{E}$, with an uncertainty band determined by the new empirical constraints \cite{HLPS2010,RW2011}. The result of such an analysis is shown in Fig.\ref{Fig16}. The condition of supporting a two-solar-mass neutron star implies that this equation-of-state must be quite stiff, with a high-density adiabatic index $\Gamma \simeq 3.7 - 3.8$. It is remarkable that, at present, the only equation-of-state that meets all constraints mentioned is the one based on standard nucleon degrees of freedom \cite{APR1998} with realistic nuclear two- and three-body forces. For a more detailed discussion of the implications of such analysis, see Ref. \cite{LP2010}.

\section{Summary and Outlook}

From the variety of existing model calculations (including those using PNJL models) one might draw the presumably premature conclusion that critical phenomena in compressed baryonic matter occur already at a density scale not much higher than that of normal nuclear matter. However, these models are so far not capable of working with the proper degrees of freedom around and below baryon chemical potentials $\sim$ 1 GeV (corresponding to quark chemical potentials around 0.3 GeV). Approaching the corresponding density scale from below, it is obvious that constraints from what we know about the nuclear matter equation of state must be seriously considered. The implementation of such nuclear matter constraints (that are well reproduced using the framework of in-medium chiral effective field theory) into a consistent picture that combines the hadronic sector of the QCD phase diagram with the high-density and high-temperature regions remains a major challenge for the future. 

Finally, in the discussion of high density, low temperature constraints on the QCD equation of state, recent new developments concerning the mass-radius relation deduced from observations of neutron stars in binaries are of great interest in reducing the set of acceptable equations of state. In particular, too soft exotic equations of state are most likely ruled out by the existence of a two-solar-mass neutron star, whereas an equation of state with ``standard" nuclear physics degrees of freedom and realistic correlations including three-body forces appears to be consistent
with observations.

\section*{Acknowledgements}
Sincere thanks go to my collaborators Nino Bratovic, Salvatore Fiorilla, Thomas Hell, Jeremy Holt, Norbert Kaiser, Bertram Klein and Bernhard R\"ottgers,  whose works have contributed substantially to this report. This work was partially supported by grants from BMBF, GSI and by the DFG Cluster of Excellence ``Origin and Structure of the Universe".


\begin{thebibliography}{99}
\itemsep -2pt 

\bibitem{Ch2008}
M. Cheng et al., \Journal{\PRD}{77}{014511} {2008};   \Journal{\PRD}{81}{054510} {2010}.

\bibitem{Bo2010}
S. Borsanyi et al., arXiv:1005.3508 [hep-lat] .

\bibitem{BP2010}
A. Bazanov, and P. Petreczky,  {\it PoS LAT2009}, (2009) 163;  arXiv:1005.1131 [hep-lat] .

\bibitem{Aoki2006}
Y. Aoki, Z. Fodor, S.D. Katz, and K.K. Szabo, \Journal{\PLB} {643}{46} {2006}.

\bibitem{Al2008}
M.G. Alford, A. Schmitt, K. Rajagopal, and T. Sch\"afer, 
 \Journal{\RMP} {80}{1455} {2008}; \\A. Schmitt, {\it Lecture Notes in Phys.} 811 (2010) 61. 
 
\bibitem{We2010}
 W. Weise, {\it Prog. Theor. Phys. Suppl.} 186 (2010) 390.
 
 \bibitem{NJL61}
Y. Nambu, and G. Jona-Lasinio,  {\it Phys. Rev.} 122 (1961) 345.

\bibitem{VW91}
U. Vogl, and W. Weise, {\it Prog. Part. Nucl. Phys.} 27 (1961) 195.

\bibitem{HK94}
T. Hatsuda, and T. Kunihiro,  \Journal{\PREP}{247}{221} {1994}.

\bibitem{Fu2003}
K. Fukushima, \Journal{\PRD}{68}{045004}{2003}; \Journal{\PLB}{591}{277} {2004}. 

\bibitem{RTW2006}
C. Ratti, M. Thaler, and W. Weise, \Journal{\PRD}{73}{014019}{2006}.

\bibitem{RRW2007}
S. R\"ossner, C. Ratti, and W. Weise, \Journal{\PRD}{75}{034007}{2007} .

\bibitem{HRCW2009}
T. Hell, S. R\"ossner, M. Cristoforetti, and W. Weise, \Journal{\PRD}{79}{014022}{2009}.

\bibitem{HRCW2010}
T. Hell, S. R\"ossner, M. Cristoforetti, and W. Weise, \Journal{\PRD}{81}{074034}{2010}.

 \bibitem{HKW2011}
 T. Hell, K. Kashiwa, and W. Weise, \Journal{\PRD}{83}{114008} {2011}.

\bibitem{Ko2010}
K.-I. Kondo, \Journal{\PRD}{82}{065024} {2010}.

\bibitem{MP08}
F. Marhauser and J.M. Pawlowski, arXiv:0812.1144 [hep-th].

\bibitem{Bo2003}
P.O. Bowman et al., {\it Nucl. Phys. Proc. Suppl.} 119 (2003) 323.

\bibitem{NMO2010}
T.Z. Nakano, K. Miura, and A. Ohnishi, \Journal{\PRD}{83}{016014} {2011}.

\bibitem{CHKW2010}
M. Cristoforetti, T. Hell, B. Klein, and W. Weise, \Journal{\PRD}{81}{114017} {2010}.

\bibitem{HLP2008}
Y. Hidaka, L. McLerran, and R.D. Pisarski, \Journal{\NPA}{808}{117} {2008}; \\
L. McLerran, and R.D. Pisarski, \Journal{\NPA}{796}{83} {2007}.

\bibitem{An2010}
A. Andronic et al., \Journal{NPA}{837}{65}{2010}.

\bibitem{FP2008}
Ph. de Forcrand, and O. Philipsen, {\it PoS LAT2008} (2008) 208;  arXiv:0811.3858 [hep-lat].

\bibitem{Fu2008}
K. Fukushima, \Journal{\PRD}{77}{114028}{2008}.

\bibitem{YTHB2007}
N. Yamamoto, M. Tachibana, T. Hatsuda and G. Baym, \Journal{\PRD}{76}{074001}{2007}.

\bibitem{ABHY2010}
H. Abuki, G. Baym, T. Hatsuda and N. Yamamoto, \Journal{\PRD}{81}{125010}{2010}.

\bibitem{Braun2011}
J. Braun, L.M. Haas, F. Marhauser, and J.M. Pawlowski, \Journal{\PRL}{106}{022002} {2011}.

\bibitem{KHW2011}
K. Kashiwa, T. Hell, and W. Weise, \Journal{\PRD}{84}{056010} {2011}.

\bibitem{PQM}
B.J. Sch\"afer, J.M. Pawlowski, and J. Wambach, \Journal{\PRD}{76}{074023} {2007}.

\bibitem{HPS2011}
T.K. Herbst, J. Pawlowski, and B.-J. Sch\"afer, \Journal{\PLB}{696}{58} {2011}.

\bibitem{FLM2011}
C.S. Fischer, J. L\"ucker, and J.A. M\"uller , \Journal{\PLB}{702}{438} {2011}.

\bibitem{LFA2000}
M.F.M. Lutz, B. Friman, and Ch. Appel, \Journal{\PLB}{474}{7} {2000}.

\bibitem{KFW2002}
N. Kaiser, S. Fritsch, and W. Weise, \Journal{\NPA}{697}{255}{2002};\\
S. Fritsch, N. Kaiser, and W. Weise, \Journal{\PLB}{545}{73}{2002}.

\bibitem{FKW2005}
S. Fritsch, N. Kaiser, and W. Weise, \Journal{\NPA}{750}{259}{2005}.

\bibitem{FKW2011}
S. Fiorilla, N. Kaiser, and W. Weise, arXiv:1111.3688 [nucl-th].

\bibitem{FKVW2006}
P. Finelli, N. Kaiser, D. Vretenar, and W. Weise, \Journal{\NPA}{770}{1}{2006}.

\bibitem{HKW2009}
J.W. Holt, N. Kaiser, and W. Weise, \Journal{\PRC}{79}{054331}{2009},\\ \Journal{\PRC}{81}{024002}{2010}.

\bibitem{HKW2011a}
J.W. Holt, N. Kaiser, and W. Weise, \Journal{\NPA}{870}{1}{2011},\\ \Journal{\EPJA}{47}{128}{2011}.

\bibitem{KHW2008}
N. Kaiser, P. de Homont, and W. Weise, \Journal{\PRD}{77}{025204}{2008}.

\bibitem{FKW2011a}
S. Fiorilla, N. Kaiser, and W. Weise, arXiv:1104.2891 [nucl-th] .

\bibitem{GL89}
P. Gerber and H. Leutwyler, \Journal{\NPB}{321}{387}{1989}.

\bibitem{Kai99}
N. Kaiser, \Journal{\PRC}{59}{2945}{1999}.

\bibitem{LP2007}
J.M. Lattimer, and M. Prakash, {\it Phys. Reports} 442 (2007) 109 .

\bibitem{APR1998}
A. Akmal, V.R. Pandharipande, and D.G. Ravenhall, \Journal{\PRC}{58}{1804}{1998}.

\bibitem{Sch2010}
A. Schmidt, {\it Lecture Notes in Phys.} 811 (2010) 1 .

\bibitem{Dem2010}
P.B. Demorest, T. Pennucci, S.M. Ransom, M.S.E. Roberts, and J.W.T. Hessels, \\{\it Nature} 467 (2010) 1081 .

 \bibitem{SLB2010}
A.W. Steiner, J. Lattimer, and E.F. Brown, {\it Astrophys. J.} 722 (2010) 33 .

\bibitem{HLPS2010}
K. Hebeler, J.M. Lattimer, C.J. Pethick, and A. Schwenk, \Journal{\PRL}{105}{161102}{2010}.

\bibitem{RW2011}
B. R\"ottgers, {\it Bachelor Thesis}, TU Munich (2011) .

\bibitem{LP2010}
J.M. Lattimer, and M. Prakash, arXiv:1012.3208 [astro-ph.SR] ;\\ J.M. Lattimer, {\it Prog. Theor. Phys. Suppl.} 186 (2010) 1 .



\end{thebibliography}
\end{document}